\documentclass{aa}  
\usepackage{graphicx}
\usepackage{txfonts}
\usepackage{float}
\begin{document} 

   \title{Solar photospheric spectrum microvariability }

   \subtitle{II. Observed relations to magnetic activity and radial-velocity modulation}
  
   \author{Dainis Dravins
          \inst{1} and
       Hans-G\"{u}nter Ludwig
           \inst{2}
}
%
%
\institute{Lund Observatory, Division of Astrophysics, Department of Physics, Lund University, SE-22100 Lund, Sweden\\
              \email{dainis@astro.lu.se}
\and
     Zentrum f\"{u}r Astronomie der Universit\"{a}t Heidelberg, Landessternwarte, K\"{o}nigstuhl 12, DE--69117 Heidelberg, Germany\\
              \email{hludwig@lsw.uni-heidelberg.de}
             }
 
\date{Received February 23, 2024; accepted April 7, 2024} 

 
\abstract
   { Searches for small exoplanets around solar-type stars are limited by stellar physical variability, such as jittering of the apparent photospheric radial velocity.  While chromospheric variability is well studied, observing, modeling. and understanding the much smaller fluctuations in photospheric spectral line strengths, shapes, and shifts is challenging. }
   {Extreme precision radial-velocity spectrometers enable extreme precision stellar spectroscopy and time series of the Sun seen as a star permit monitoring of its photospheric variability.  To understand such microvariability through hydrodynamic 3D models will require diagnostics from different categories of well-defined photospheric lines with specific formation conditions.  Fluctuations in their line strengths may well correlate with radial-velocity excursions and identify observable proxies for their monitoring. }
   {From three years of HARPS-N observations of the Sun-as-a-star at $\lambda$/$\Delta\lambda$ $\sim$100,000, one thousand low-noise spectra are selected, and line absorption measured in \ion{Fe}{i}, \ion{Fe}{ii}, \ion{Mg}{i}, \ion{Mn}{i}, H$\alpha$, H$\beta$, H$\gamma$, \ion{Na}{i}, and the {\it{G}}-band.  Their variations and likely atmospheric origins are examined, also with respect to simultaneously measured chromospheric emission and apparent radial velocity. }
   {Systematic line-strength variability is seen, largely shadowing the solar-cycle evolution of \ion{Ca}{ii}~H\,\&\,K emission, but with smaller amounts, typically on a sub-percent level.  Among iron lines, greatest amplitudes are for \ion{Fe}{ii} in the blue, while the trends change sign among differently strong lines in the green \ion{Mg}{i} triplet and between Balmer lines.  Variations in the {\it{G}}-band core are greater than of the full {\it{G}}-band, in line with theoretical predictions. No variation is detected in the semi-forbidden \ion{Mg}{i} $\lambda$\,457.1 nm.  Hyperfine split \ion{Mn}{i} behaves largely similar to \ion{Fe}{i}.  For lines at longer wavelengths, telluric absorption limits the achievable precision.  }
   {Microvariability in the solar photospheric spectrum displays systematic signatures among various features.  These thus measure something different than the classical \ion{Ca}{ii}~H\,\&\,K index while still reflecting a strong influence from magnetic regions.   Although unprecedented precision can be achieved from radial-velocity spectrometers, current resolutions are not adequate to reveal changes in detailed line shapes, and their photometric calibration is not perfect.  A forthcoming priority will be to model microvariability in solar magnetic regions, which could also provide desired specifications for future instrumentation toward exoEarth detections. }


\titlerunning{Solar spectrum microvariability. II}
\authorrunning{D. Dravins \& H.-G.Ludwig}
\maketitle

\section{Introduction}

Extreme precision radial-velocity spectrometers designed for a wavelength stability corresponding to Doppler shifts of $\sim$1~m\,s$^{-1}$ or better are now operating at several telescopes, and are planned also at future facilities, as summarized in Paper~I \citep{dravinsludwig23}.  Their main task is to search for periodic modulation of stellar radial velocities induced by orbiting exoplanets.  The detection of planets with successively smaller mass requires to recognize successively smaller velocity amplitudes: for an Earth-mass planet orbiting a solar-mass star in a one-year orbit, the expected signal amounts to at most only 0.1 m\,s$^{-1}$ \citep[e.g.,][]{halletal18}.  Although current instrumental precisions actually begin to approach these levels, radial-velocity fluctuations of much greater amplitude are contributed by stellar phenomena such as convective motions, oscillations, or the presence of dark starspots or bright magnetic structures.  

While elaborate analyses of observed time series now succeed in eliminating much of the non-planetary signals, even the currently best modeling is unable to extract planetary signatures with amplitudes much below 1 m\,s$^{-1}$.  The limitations are thus no longer instrumental but lie in understanding the complexities of stellar radiation and spectral line formation, manifest as a spectral jitter of the apparent radial velocity and a photometric flickering of the irradiance.  The need to mitigate such effects on radial-velocity signatures of exoplanets was realized already long ago but with current instrumental precisions approaching the levels for finding Earth-mass exoplanets in the habitable zones around solar-type stars, stellar microvariability has become the critically limiting factor toward reaching 0.1~m\,s$^{-1}$ or better \citep{crassetal21, fischeretal16}.

One step toward the challenging goal of finding an exoEarth, would be to identify some proxy for the excursions in apparent radial velocity.  Such a quantity should be possible to measure from the ground independently from the radial velocity as a whole.  In Paper~I, time sequences of synthetic spectra computed from 3D hydrodynamical models of the nonmagnetic solar photosphere were scrutinized to identify such parameters.  It was found that while most spectral lines fluctuate in phase, the precise amplitudes differ by up to about one tenth of their values and depend on the spectral line strength, ionization state, excitation level, and wavelength region.  Although the differential effects are small, sufficiently precise observations should enable to identify and compensate for such short-term influences from surface convection.  However, the spectrum of the full Sun comprises contributions from also various magnetic structures which produce long-term spectral modulation during the solar activity cycle.  Variations in strong chromospheric lines from \ion{Ca}{ii} and other species have been extensively monitored in the past but their correlation with jittering of radial velocities cannot be exact since those are determined from photospheric spectra only.  The present paper examines solar spectral microvariability from an observational side, both to identify variations in different categories of photospheric spectral lines, but also to better understand what the practical precision limits are in data from current types of instruments.

\section{Microvariability in integrated sunlight}

The spectrum of the Sun seen as a star fluctuates on different timescales and across various wavelength regions but the bulk of the energy output originates from the photosphere and radiates in the optical. This visual portion is rich in lines and is the one primarily used to measure radial velocities in solar-type stars.  Here, the variations are modest: fluctuations of the total solar irradiance can be followed from space but are barely discernible from the ground.  Significant variations in the optical spectrum are limited to strong lines or line components influenced by the chromosphere or corona, such as the central emissions in \ion{Ca}{ii}~H\,\&\,K, the strongest Balmer lines from hydrogen, the infrared \ion{Ca}{ii} triplet or the \ion{He}{i} $\lambda$\,1083 nm feature.  Full-disk variability in such lines can often be traced back to the appearance of solar surface plages and other structures as seen on spectroheliograms in the same wavelengths.  In the context of the 11-year sunspot cycles, chromospheric emission in especially the \ion{Ca}{ii} H\,\&\,K lines has been monitored since more than a century \citep{chatzistergos22, singhetal21}.  An activity measure from the \ion{Ca}{ii}~H\,\&\,K emission is often quantified as the Mt.Wilson S-index or, for different spectral types, with the closely related measure $R\,'\rm_{HK}$ used to monitor stellar activity cycles \citep{egelandetal17}. The S-index is based on the summed relative strengths in bandpasses centered on the \ion{Ca}{ii}~H\,\&\,K  line cores \citep{vaughanetal78}.  However, variability in ordinary Fraunhofer lines is much more elusive.  To detect, understand, and exploit any variability in such lines, high-fidelity observations at high spectral resolution are required with very stable instrumentation.

\subsection{Early searches for radial-velocity fluctuations}

At Kitt Peak National Observatory, a solar line-strength monitoring program was started around 1974.  As an outgrowth of that program, line-shape differences in \ion{Fe}{i} lines were identified as different bisector shapes and shifts between regions of magnetic and nonmagnetic granulation.  Using the then new Fourier Transform Spectrometer (FTS), a diminished convective blueshift over magnetic areas could be observed, and a corresponding signature seen in integrated sunlight between different years of the solar activity cycle \citep{livingston82, livingston83, livingston84}.  Similar trends were confirmed in spectra near the solar-disk center \citep{cavallinietal86} while the systematic shrinkage of the bisector amplitude when going from quiet to magnetically active granulation was documented by, e.g., \citet{brandtsolanki90}, \citet{cavallinietal85, cavallinietal88}, and \citet{immerschittschroeter89}.  

The disturbing effects of an 11-year activity-cycle modulation of solar wavelengths on the then concurrent searches for exoplanets were understood, especially since -- at that time -- the searches were mainly for objects comparable to Jupiter in its 12-year orbit.  A need was thus realized to understand and mitigate such effects to enable exoplanet detection from their radial-velocity signatures \citep[e.g.,][]{demingetal87, dravins85, dravins89, wallaceetal88}.  The FTS at Kitt Peak enabled precise wavelength determinations and was used by \citet{demingetal87} and \citet{demingplymate94} to follow some infrared lines around 2.3 $\mu$m in integrated sunlight over several years, indicating an activity-cycle modulation, although a null result was found from other observations of moonlight at lower spectral resolution in the violet \citep{mcmillanetal93}.

\subsection{Kitt Peak monitoring of the spectrum of sunlight}
 
The most comprehensive search for variations in the visual spectrum of the Sun seen as a star was then carried out at Kitt Peak, spanning some 35 years.  Observations were made about once per month, using the double-pass spectrograph at the then McMath (the later McMath-Pierce) solar telescope, with its light input modified to approximate integrated sunlight.  These observations from 1974-2009 are summarized by \citet{livingstonetal07, livingstonetal10}, with details in  \citet{livingstonholweger82}, \citet{livingstonwallace87}, \citet{livingstonetal77, livingstonetal11}, and \citet{whitelivingston78, whitelivingston81}. 

Their results show that various \ion{Ca}{ii}~K features track the 11-year magnetic cycle based on sunspot number with a peak amplitude in central intensity of $\sim$37\%. The wavelength of the mid-line core absorption feature, called K$_3$, as referenced to nearby photospheric \ion{Fe}{i}, displays an activity cycle variation with a full-disk amplitude of 0.3 pm (3 m{\AA}); 0.6 pm at disk center. Other chromospherically influenced lines, such as \ion{He}{i} $\lambda$\,1083 nm, H$\alpha$, and the CN $\lambda$\,388 nm bandhead, also track the \ion{Ca}{ii}~K intensity, although with smaller amplitudes.  The core of the \ion{Ca}{ii} $\lambda$\,854.2 nm line shows solar-cycle changes \citep{pietarilalivingston11} while measurements with other instrumentation show the \ion{He}{i} $\lambda$\,1083 nm changing its equivalent width \citep{harvey84, harveylivingston94}.  Further lines monitored included \ion{C}{i} $\lambda$\,538 nm, cores of stronger \ion{Fe}{i}, \ion{Na}{i}~D$_1$ and D$_2$, and \ion{Mg}{i}~b, but with less clear conclusions about possible variations in their strengths. 

Although \ion{Fe}{i} bisector amplitudes were observed to vary over the solar cycle, they appeared not to be in phase with other activity indices \citep{livingstonetal99}.  For bisector shapes in integrated sunlight, \citet{bruninglabonte85} found no correlation with solar magnetic flux for the same day, but instead with a 30-day average, indicating that the line asymmetries relate to extended areas of magnetic plage rather than to current sunspot regions. Similar conclusions are drawn from recent radial-velocity spectrometer data, where a passage of a sunspot group induces a radial-velocity change, correlating with line asymmetry modulations, but leading those by some 3 days \citep{colliercameronetal19}.   

\subsection{Detection limits for photospheric changes}

While the solar-cycle modulation of chromospheric indices reaches amplitudes of several few tens of per cent, the full-disk variability in the strength of photospheric lines is much smaller, generally below one percent \citep{livingstonetal07, mitchelllivingston91}, and with discrepant or inconclusive results reported between different solar activity periods.  At these accuracy levels, challenges emerge in keeping instrumentation and observational conditions stable over longer epochs.  The only photospheric line for which an apparently significant change was reported from the Kitt Peak monitoring program was \ion{Mn}{i} $\lambda$\,539.47 nm, with its variability tightly correlated with the \ion{Ca}{ii}\,K$_3$ intensity.  Such a correlation would suggest the variability to be related to magnetic plage regions but the authors found arguments against such a connection \citep{livingstonwallace87}.

\citet{livingstonholweger82} and \citet{livingstonwallace87} provide extensive discussions on observational limits for the Kitt Peak program, and potential error sources.  Their spectrometer -- thanks to its double-pass design -- had a very clean instrumental profile at a (single-pass) spectral resolution of $\sim$\,60,000, but being in air (not vacuum), it was affected by internal seeing and, being a scanning instrument, also by scintillation.  Some types of variation were identified as likely due to spectrograph alignments, e.g., thermal drift introduces a time-dependent asymmetric instrumental profile.  A change of diffraction gratings somewhat modified the mode of integrating over the solar disk.  Slight changes occurred from the recollimation required between observations in visual and infrared.  The central depths of absorption lines would be valuable to monitor but they directly depend on the instrumental profile and their recorded variations showed a component which mimics spectrograph defocusing.  Kitt Peak is affected by alternating wet and dry atmospheric conditions, and corrections for telluric water vapor absorption have to be made.  Despite careful observational work, instrumental systematics on the desired fidelity levels are difficult to fully avoid and, following some recalibrations, some of the earlier variability indications of the weak \ion{C}{i} $\lambda$\,538 nm line, originally suggesting solar temperature changes, were acknowledged as ``no longer valid'' \citep{livingstonetal07}.  Probably, these types of observations stretch the accuracy limits for spectrographs that are operated in air, and have to be readjusted by successive observers for their use in different observing programs.  To conclusively identify solar photospheric spectrum variability appears to require vacuum instruments of the extreme precision type developed for exoplanet searches.  

\subsection{Space-based observations}

While spaceborne high-resolution optical spectrometers still lie in the future \citep{plavchanetal19}, solar radiation is being monitored from space using various radiometers with  a certain spectral resolution.  From near-daily measurements \citet{criscuolietal23, marchenkodeland14}, and \citet{marchenkoetal21} find that activity indices of the hydrogen Balmer lines, computed as core-to-wing ratios, show variability on solar rotational timescales, largely following that of the total solar irradiance and thus following photospheric rather than chromospheric indices.  Irradiance variations in various spectral segments can be followed throughout the optical spectrum although the spectral resolution is modest as compared to ground-based spectrometers.  From modeling the effects of magnetic fields, one expects the maximum of the spectral brightness variability on timescales greater than a day to occur around the CN bandhead between 380-390 nm \citep{shapiroetal15}.

\begin{figure}
 \centering
 \includegraphics[width=\hsize]{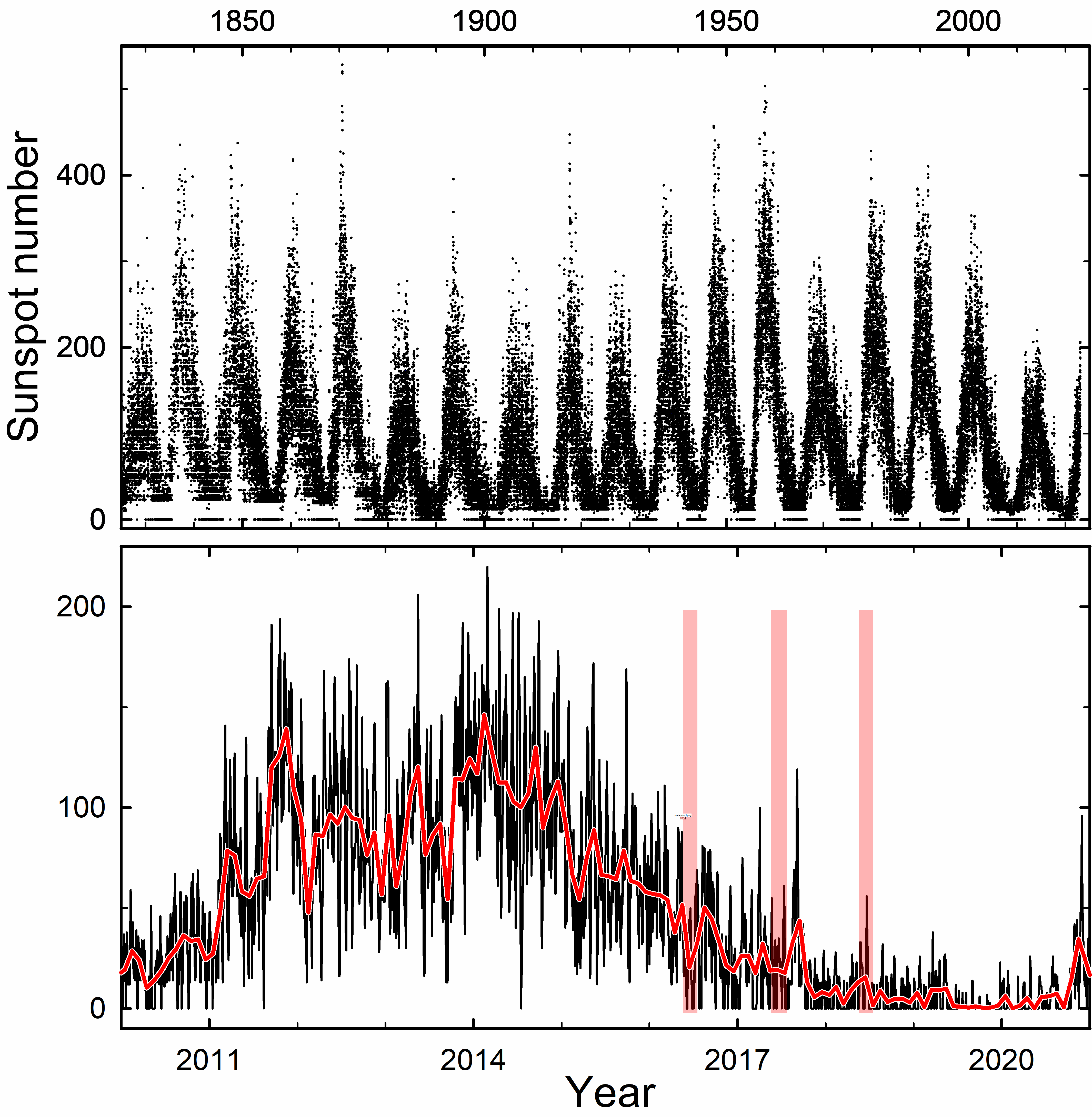}
     \caption{Top: A 200-year span of sunspot numbers places current observations in a wider perspective. Around activity minima, days without reported sunspots are not unusual. Bottom: The recent Cycle 24 with daily sunspot numbers from WDC-SILSO (World Data Center - Sunspot Index and Long-term Solar Observations, Royal Observatory of Belgium, Brussels).  Superposed red curve shows monthly averages. The selected periods from which HARPS-N data were analyzed are marked with vertical bars. }  
\label{fig:activitycycles}
\end{figure}

\begin{figure}
 \centering
 \includegraphics[width=\hsize]{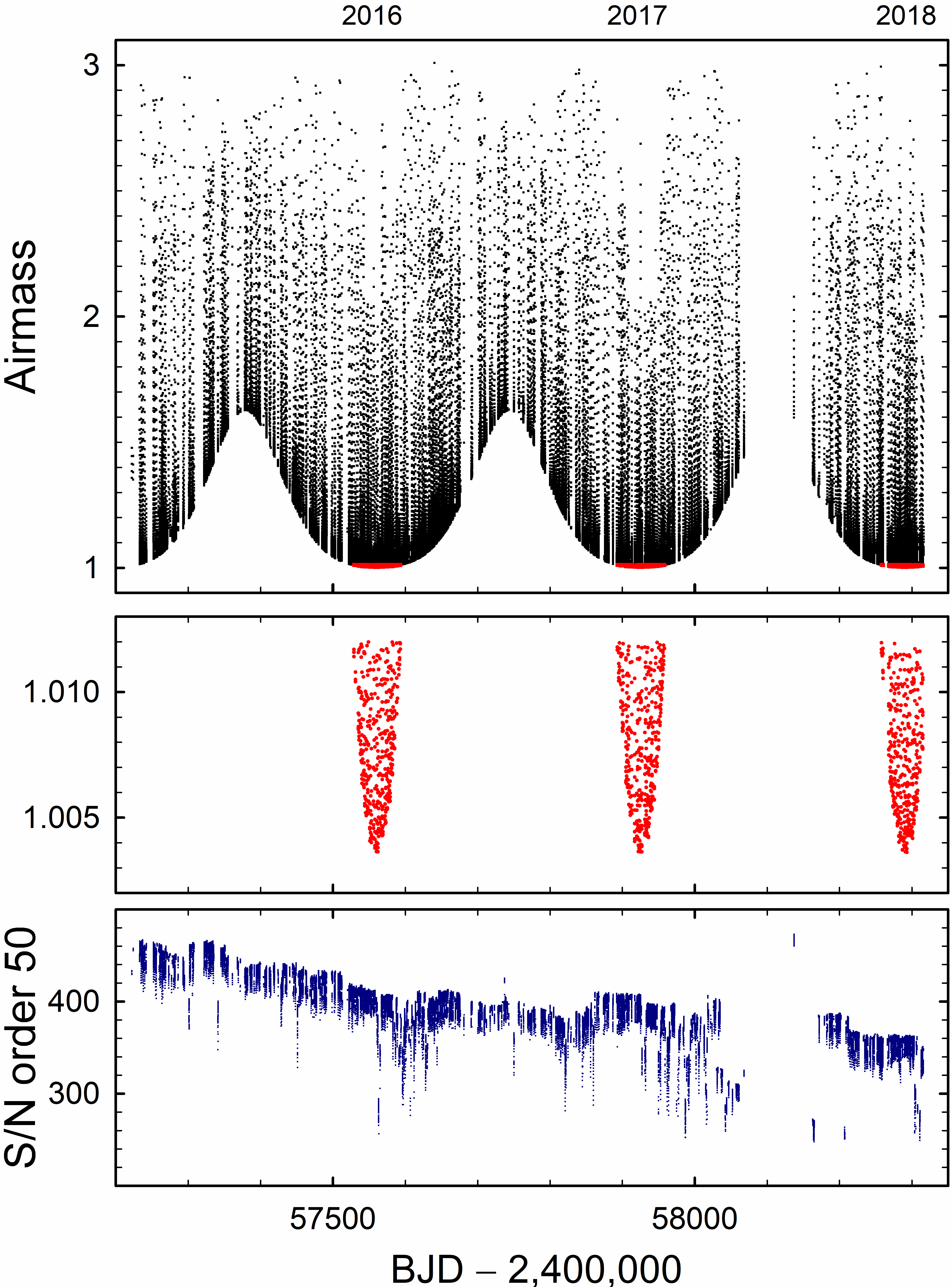}
     \caption{Top: HARPS-N observations of the Sun with each exposure marked by a point \citep{dumusqueetal21}.  Selected exposures at the smallest airmasses (red points) occurred around daily noon during the summer seasons of 2016-2017-2018.  Center: Airmasses for the 1000 selected recordings. Bottom: The signal-to-noise ratio remained very high but decreased slightly over time, apparently due to diminishing telescope and instrument transmission. } 
\label{fig:metadata}
\end{figure}

\begin{figure}
 \centering
 \includegraphics[width=\hsize]{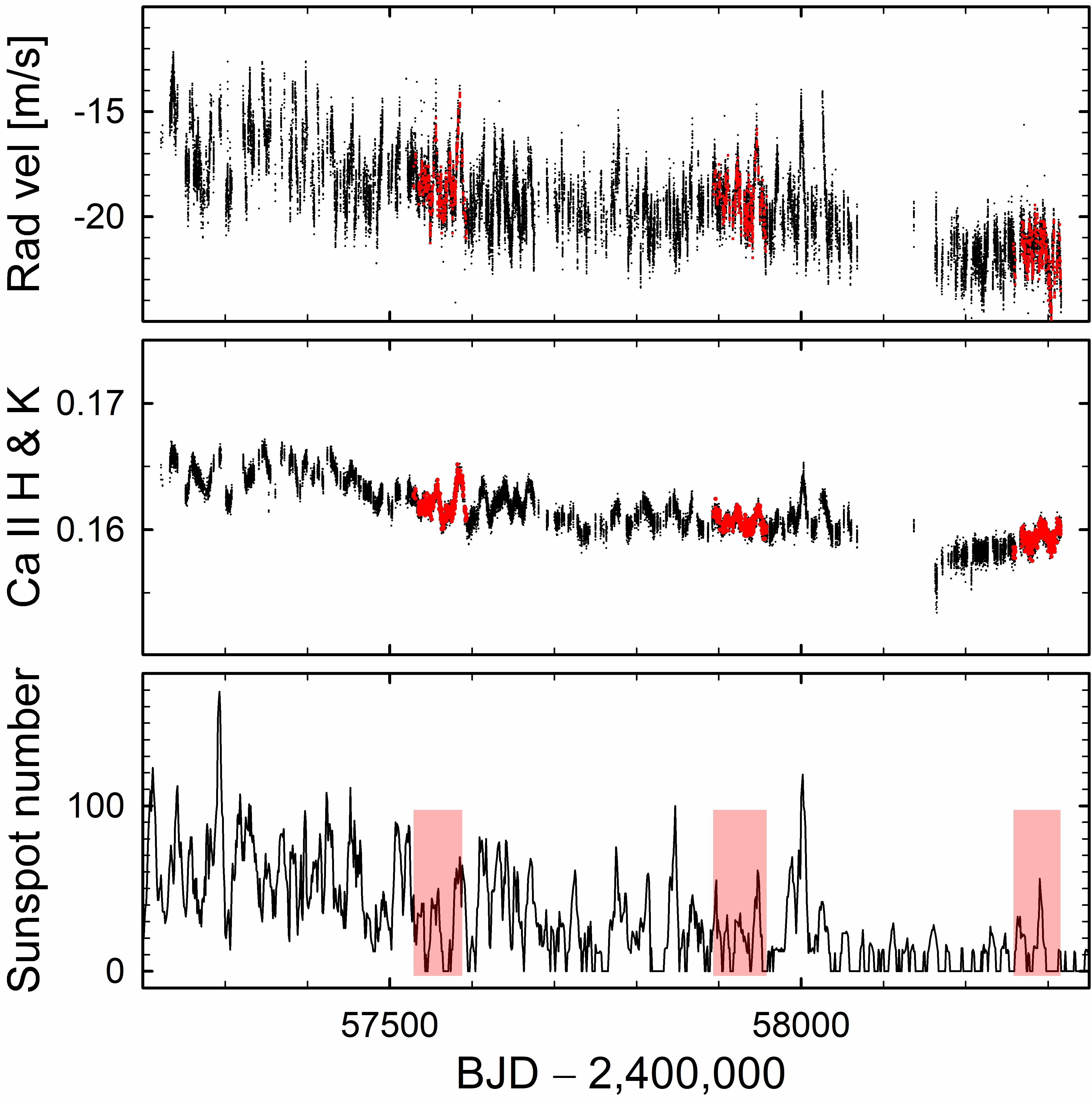}
     \caption{Top: Apparent solar radial velocity grows more negative with time (given as Barycentric Julian Date), apparently due to less magnetic granulation and a smaller suppression of convective blueshift.  Middle: The Mt.Wilson Ca II~H\,\&\,K S-index reflects the slowly declining activity level in the solar cycle, modulated by solar rotation.  Bottom: Daily sunspot numbers from WDC-SILSO. The periods from which HARPS-N data were analyzed are marked in red or shaded in color. }  
\label{fig:solaractivity}
\end{figure}

\begin{figure*}
\centering
 \includegraphics[width=18cm]{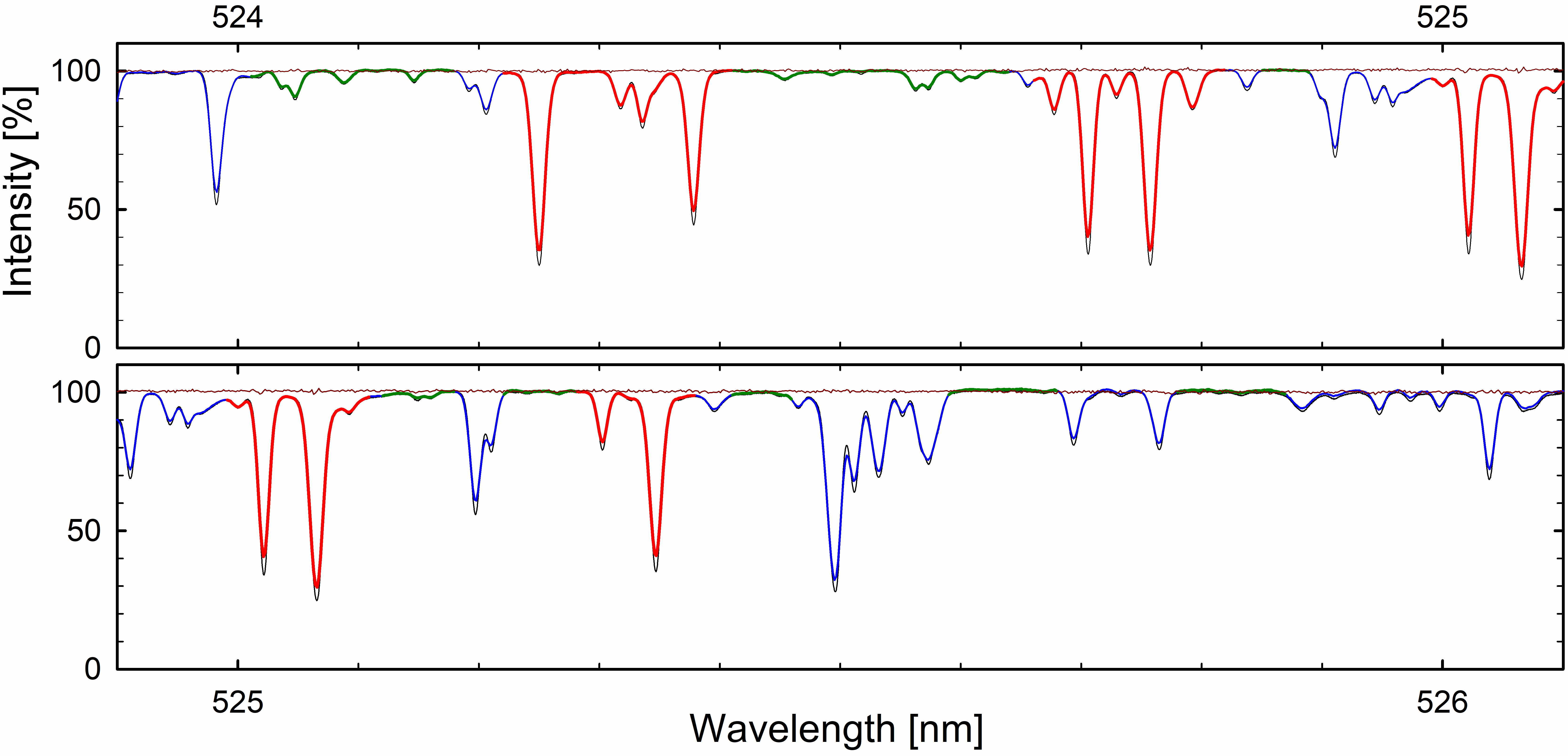}
     \caption{Example of HARPS-N spectra and selection of regions for measurement.   Background thin black line: G{\"o}ttingen solar flux atlas \citep{reinersetal16}.  Its higher spectral resolution is evident from its deeper line profiles. Blue line: Average over 100 HARPS-N exposures.  Red line: Selected regions of  \ion{Fe}{i} line absorption.  Green line: Selected reference regions of quasi-continua.  Thin brown line around 100\%: ratio between one representative HARPS-N exposure and a 100-exposure average, illustrating the random-noise level for a single exposure with nominal S/N in grating order 50 equal to 430.  The wavelength scale represents values in air (vacuum values in the Göttingen atlas were converted to those in standard air).  }  
\label{fig:520nm_general}
\end{figure*}

\subsection{Potential of radial-velocity spectrometers}

Extreme precision radial-velocity spectrometers now enable extreme precision stellar spectroscopy.  Compared to earlier types of instruments, a series of enhancements avoid several limitations, in particular: a uniformly illuminated entrance aperture defined by a scrambled optical fiber (rather than a physical slit); the optics sealed in vacuum (rather than air), an active (rather than passive) thermal control, no moving optical parts (thus no readjustment between observations), and a single-use setup for all observations and their data reduction.  Several such instruments are measuring also integrated sunlight \citep{claudietal20, dumusqueetal15, eso18, eso21, linetal22, phillipsetal16, rubenzahletal23}, and the measured velocities show consistent values among the different instruments of HARPS, HARPS-N, EXPRES, and NEID \citep{zhaoetal23}.  With such instrumentation, the prospects look promising to search for subtle microvariability in photospheric absorption lines, which previously could not be conclusively detected.   

Still, data from radial-velocity spectrometers have certain limitations.  Although their spectral resolution is termed as high, it is much lower than what can be obtained from synthetic spectra or in spectrometers at large solar telescopes and some spectral-line signatures are smeared out.  The wavelength calibration is very stable but is normally not absolute between different grating orders. The photometric precision can be high but varies among differently deep exposed spectral orders and does not reach what is feasible in dedicated photometry of small spectral portions in solar spectrometers.  As a consequence, attempts to detect variations of parts in a thousand, may require averaging several lines of physically similar parameters.  This applies especially for narrow features, covering fewer detector pixels. 

It is difficult to tell how far the precision can be pushed but one may have to evaluate how the instrumental profile shapes, widths and asymmetries vary across the focal plane and depend on not only the spectrometer optics, but also on, e.g., the physical segmenting and stitching inside the CCD detector, and on its electronic readout direction, as documented for the HARPS spectrometer by \citet{locurtoetal12a, locurtoetal12b}, \citet{milakovicjetwha23}, \citet{molaroetal13}, and \citet{zhaoetal14, zhaoetal21}.  A detailed evaluation of the calibration and correction of numerous subtle effects in the HARPS instrument is by \citet{cretignieretal21}.

Like all observations from Earth, the spectra are contaminated by superposed telluric absorption lines.  Those from water vapor are especially variable in strength, reflecting both local meteorological conditions and the varying airmass through which the Sun is observed.  Further, their exact wavelength positions relative to the solar spectrum change because of Doppler shifts induced by both the Earth's daily rotation and its annual motion in our somewhat eccentric orbit around the Sun.  
In proposing space-based radial-velocity instrumentation, \citet{plavchanetal19} argue that telluric absorption will limit radial-velocity precisions to $\sim$0.1 m\,s$^{-1}$ at wavelengths beyond $\sim$700 nm (and becoming worse in the infrared).  A detailed analysis of how measured parameters of photospheric lines can be affected by telluric contamination when observed through different airmasses is by \citet{vincevince10}.  

With adequate effort, telluric effects can be minimized to a certain level \citep[e.g.,][]{allartetal22, bakeretal20, cunhaetal14, ivanovaetal23, kjaersgaardetal23, xuesongwangetal22} but, unless very carefully verified, it might introduce additional uncertainties among very numerous spectra recorded under somewhat variable atmospheric conditions.  

A study of photospheric spectrum microvariability as measured by extreme precision radial velocity spectrometers has the additional advantage of having contemporaneous radial-velocity values, as computed by a cross-correlation for the spectrum as a whole, as well as various statistical parameters for the observed spectrum.  This enables searches for possible relations between measured variability parameters and modulation of apparent radial velocity.  Values for the radial velocities are obtained by cross-correlating (in the wavelength domain) the intensities of entire measured spectra against a synthetic digital mask representing a solar-type spectrum, with weighted contributions from different spectral lines \citep{baranneetal96}. 

\section{HARPS-N observations of the Sun as a star}

Building upon the experience from the past Kitt Peak program, evaluating the theoretical microvariability studies in Paper~I \citep{dravinsludwig23}, and taking advantage of current instrumentation developments, a search for microvariability in the solar photospheric spectrum was undertaken. On La Palma, a small sunlight-integrating telescope feeding the HARPS-N\footnote{High Accuracy Radial velocity Planet Searcher for the Northern hemisphere} spectrometer on the TNG\footnote{The 3.5 m Telescopio Nazionale Galileo} telescope building, has kept observing the Sun as a star since the summer of 2015 \citep{cosentinoetal12, phillipsetal16, thompsonetal20}.  Its first public data release covered the three-year period from July 2015 through July 2018 (Fig.\ \ref{fig:activitycycles}), a declining phase of the solar activity cycle \citep{colliercameronetal19, dumusqueetal15, dumusqueetal21,maldonadoetal19, milbourneetal19}.  These data comprise 34,550 spectra (about 65\% of all observations from that period, selected for best quality), recorded with exposure times of 300 seconds (chosen to largely average out the 5-minute part of the p-mode oscillations), with a usual cadence between exposures of $\sim$5.4 minutes.  The reduced spectra are corrected for the spectrometer blaze function, while the values for the radial velocity are those for the Sun in the heliocentric (rather than solar-system barycentric) rest frame. The flux is provided per pixel, in a stepsize of 0.82~km\,s$^{-1}$, with 3.2 pixels per spectral resolution element of $\lambda$/$\Delta\lambda$\,$\sim$115,000, covering the wavelength interval of 387-691 nm with $\sim$210,000 spectral data points.  Besides the intensity spectra as such, the dataset includes additional measures such as parameters of the cross-correlation function, and chromospheric activity indicators. During this three-year period, a slight decrease in the recorded flux level (apparently due to decreasing transmission of the transparent telescope cover) contributed to a slight successive decrease of the S/N ratio.  The authors \citep{dumusqueetal21} express the hope ``that the community will use such data ... with the goal of enabling the detection of other Earths''; this work is one attempt in that direction.

\begin{figure*}
\sidecaption
 \includegraphics[width=12.9cm]{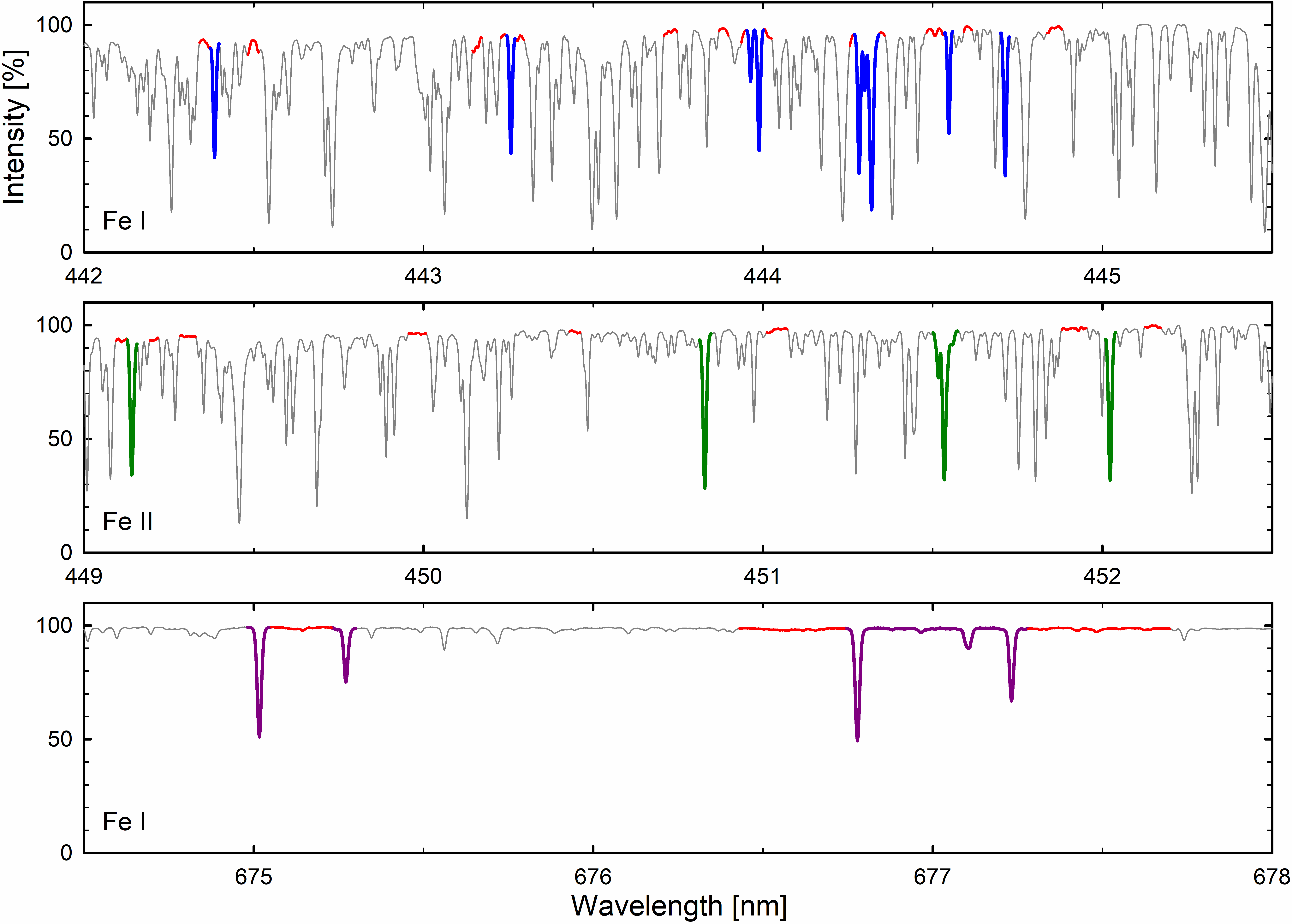}
     \caption{Example of \ion{Fe}{i} line selections in shortward parts of the spectrum (blue), of \ion{Fe}{ii} (green), and of \ion{Fe}{i} at long wavelengths (dark purple).   Absorption features are truncated at places of small intensity gradients, occasionally embracing several lines.  Intermingled pseudocontinua, whose averages are used as an intensity reference, are marked in red. }  
\label{fig:fe_line_groups}
\end{figure*}

\begin{figure}
\centering
 \includegraphics[width=\hsize]{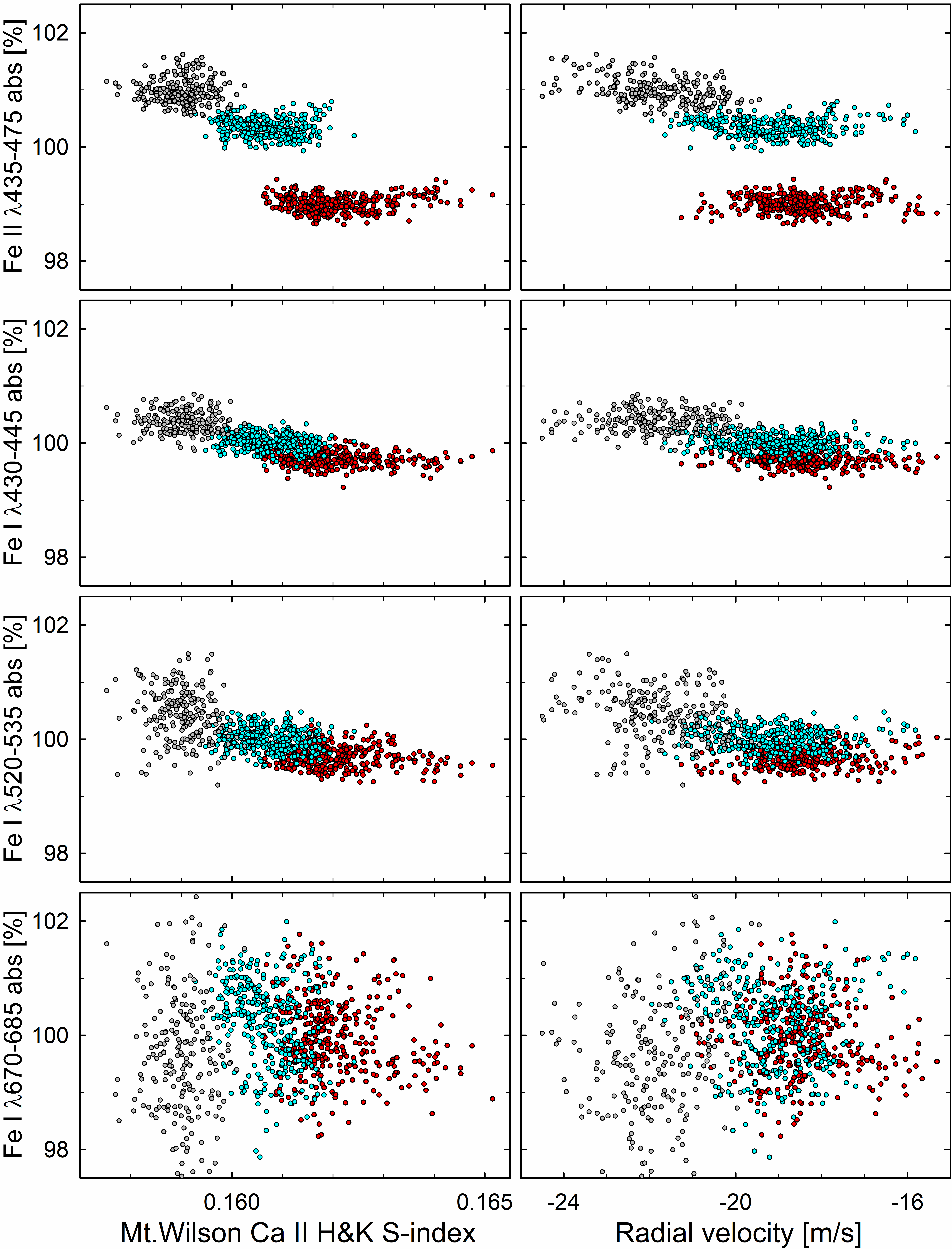}
     \caption{Relative changes of \ion{Fe}{i} and \ion{Fe}{ii} absorption line equivalent widths during the measured periods in 2016 (red), 2017 (cyan), and 2018 (gray).  Left column shows the dependence on the \ Mt.Wilson \ion{Ca}{ii}~H\,\&\,K chromospheric activity S-index; right column on the radial velocity as obtained from the full HARPS-N spectrum.  A lower S/N in the grating orders housing weaker lines at the longest wavelengths accounts for a greater spread of points in the bottom frames. The lines included in each group are listed in Table~\ref{table:linelist1}.}  
\label{fig:all_fe_lines_abs}
\end{figure}

\subsection{Previous analyses of HARPS-N full-disk solar data}

This sequence of solar data has been scrutinized by the HARPS-N team itself.  Thus, \citet{colliercameronetal19} carried out a thorough analysis using cross-correlation functions (CCF).  Sources of velocity excursions were identified, concluding that for timescales below a day, the granulation signal dominates, with a half-life of 15 minutes.  On longer time-scales, magnetic activity dominates, with active-region passages shifting the apparent radial velocity by several m\,s$^{-1}$, accompanied by spectral-line asymmetries that are shifted in time relative to the velocity signal.  Long-term trends in the approach to solar minimum appear as slowly changing parameters of the CCF, mirroring a decline in sunspot number and possibly tracing the evolving magnetic network.  With weakened network, an apparent decrease in effective temperature could result \citep{ludwigetal23, vogler05}.  Sunspot disk passages trigger peaks in the CCF FWHM while the CCF bisector signature of line asymmetry remains for longer, tracing inhibition of convection in active-region faculae.

The detectability of synthetic planetary signals in HARPS-N solar data was tested by \citet{colliercameronetal21} and \citet{langellieretal21}, suggesting that while improved models of stellar variability will be required, detailed analyses of various correlation functions should be able to segregate stellar magnetic activity.  The inverse problem, i.e., synthesizing a full-disk signature from modeled granulation was considered by \citet{ceglaetal19} and \citet{palumboetal22}.  In another study of these years of observations, \citet{maldonadoetal19} studied \ion{Ca}{ii}\,H \& K, Balmer lines, \ion{Na}{i}\,D$_1$ \& D$_2$, and \ion{He}{i}\,D$_3$, finding that activity indices defined in bandpasses around H$\alpha$, H$\beta$, and H$\gamma$ lines are anti-correlated with the \ion{Ca}{ii} S-index.  Also from HARPS-N solar spectra, \citet{thompsonetal20} further explored activity-related spectral variations.

\section{Selecting subsets of HARPS-N solar data}

Both the past Kitt Peak survey, our modeling in Paper~I, and the results by \citet{maldonadoetal19} indicate that somewhat different behavior among various classes of spectral lines is to be expected but variations are likely to be tiny, possibly at (or beyond) the detection limit.  A detailed analysis of the entire large HARPS-N data set was not seen as practical, and considerations were taken to select realistically limited subsamples of the best data available.

\subsection{Comparing HARPS-N to solar spectrum atlases}

First, a sample of 100 spectra was examined to understand more about the character of likely noise sources.  How extensive averaging of successive exposures that is feasible without hitting systematic effects depends on to what extent its noise has a random character, such that stacking of spectra decreases the noise in a stable manner.  To verify such behavior, traces of the reduced HARPS-N spectra were compared to high-fidelity spectrum atlases of integrated sunlight \citep{kuruczetal84, reinersetal16}, confirming that the spectra look very stable and that their averaging over multiple exposures indeed shows a stable convergence (see Fig.~\ref{fig:520nm_general}).  However, the continuum levels of the HARPS-N spectra are not normalized over broader spectral regions and, on some accuracy level, the comparison to spectral atlases may start to show physical differences between spectra recorded at different epochs of the solar activity cycle or at different times of year.  [Bi]annual variations of the apparent solar rotational velocity are caused by two effects: With the Earth’s orbital plane inclined relative to the solar equator, the rotating Sun is viewed from slightly different angles during the year.  The Earth’s prograde orbit is somewhat eccentric, with our orbital angular velocity greater near perihelium, when the velocity vector more closely tracks the sense of the solar rotation, and the apparent solar angular velocity thus decreases.  Although small, these effects are seen in HARPS-N data as [bi]annual modulations of spectral-line widths \citep{colliercameronetal19}. 

In the lowest-noise spectral orders, a nominal S/N ratio $\sim$400 is reached in the continuum for each data point with stepsize of 0.82~km\,s$^{-1}$  (Fig.\ \ref{fig:metadata}).  One spectral line of medium strength may cover some 25 such points, and a selection of ten similar lines, averaged over ten recordings spanning one hour, if affected by random noise only, could then in principle reach S/N values in excess of 10,000, even in absorption lines well below the continuum.  In reality, however, approaching such precision would require extremely precise control of all other noise sources.  One of the aims of the present project is to try to understand how far the search for spectral microvariability can be pushed with existing instruments before being overtaken by systematics.

\begin{figure*}
\sidecaption
 \includegraphics[width=12.9cm]{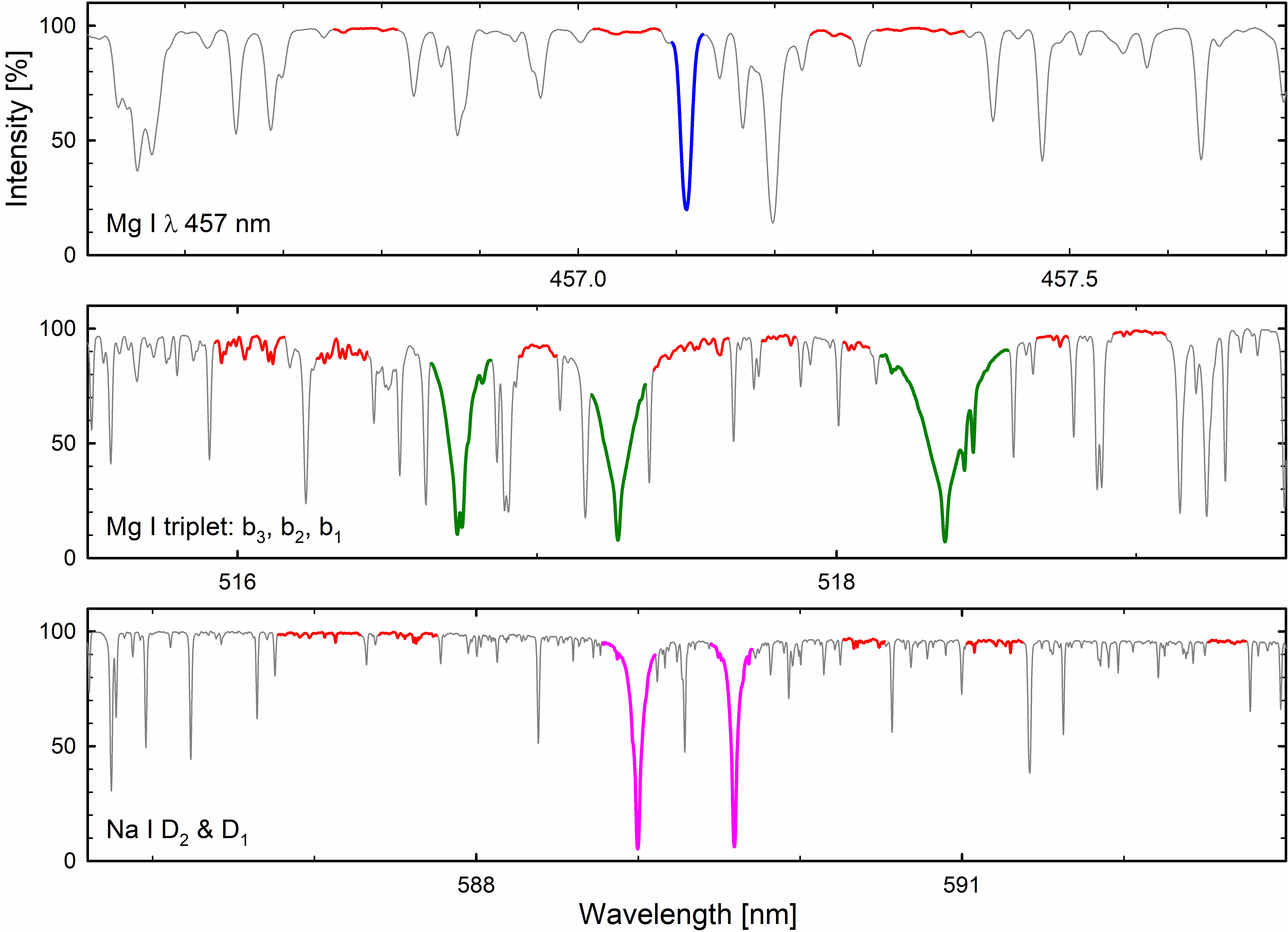}
     \caption{Top: Selected portion of the \ion{Mg}{i} 457.1 nm absorption line and its surrounding continuum reference segments (red). Center: Selection of the \ion{Mg}{i} b line triplet lines and their continuum reference segments. Bottom: \ion{Na}{i} D${_2}$ and D${_1}$ lines and their continuum segments. }  
\label{fig:chromospheric_lines}
\end{figure*}

\subsection{Criteria for selecting spectral exposures}

Since a detailed analysis of all available data was not seen as practical, some subset of spectral exposures had to be selected. Various criteria were considered: with respect to the level of solar activity, with a uniform sampling in time, of highest S/N values, and other.  It was concluded that one limiting source of non-random noise likely could be the varying amount of atmospheric telluric absorption.  Even if the strongest water vapor lines in the yellow and red spectral regions perhaps could be avoided, micro-tellurics exist throughout the spectrum, and contribute noise beyond the random photometric component.  Such effects of telluric absorption and various methods to remove them have been discussed by numerous authors, i.a., \citet{artigauetal14, cunhaetal14, ivanovaetal23, kjaersgaardetal23, lisogorskyietal19, smetteetal15, xuesongwangetal19}, and references therein, some also concluding that observations from outside the atmosphere will ultimately be required for exoEarth detection \citep{plavchanetal19}.  For the present study, in order to minimize both the amount of tellurics and their variation between exposures, it was decided to select exposures from the smallest airmasses only, implying summertime observations around daily noon (which, for the La Palma latitude, also implies close to zenith).  The full HARPS-N set of 34,550 spectra includes recordings at also large airmasses Z up to almost 3.1; $\sim$8,600 are at Z$<$1.1 and some 2,200 are below Z=1.02.  

A total of 1200 exposures with the smallest airmasses were first retrieved from the dataset.  From this group, the 1000 apparently best spectra were selected, removing those with the lowest S/N values during each observing season.   On some days, significant decreases of the S/N were noted (Fig.~\ref{fig:metadata}), which in independent meteorological records could be identified as likely due to the {\it{calima}} phenomenon, when the atmospheric transparency above La Palma is diminished by airborne dust from Sahara.  However, that should not generate sharp spectral lines (as opposed to telluric lines at larger airmasses). The final sample of 1000 spectra covers Z~=~1.0036 to 1.0120, and originates from the dates May 20 – July 18, 2016; May 19 – July 23, 2017; and May 19 – July 15, 2018.  For the selected samples, Fig.\ \ref{fig:metadata} shows the airmass distribution over time, and the corresponding S/N ratios for the {\'e}chelle order nr.\ 50, as listed in the data set.  The S/N values for the selected spectra hover around 400 during the 2016 and 2017 seasons, and around 350 in 2018. The decreasing solar activity during this period is seen as both a decrease in the Mt.Wilson \ion{Ca}{ii}~H\,\&\,K S-index, and as a gradually more negative apparent radial velocity, reflecting the smaller area coverage of plages with magnetically disturbed granulation with their locally smaller convective blueshifts \citep{meunier21}.   Fig.\ \ref{fig:solaractivity} shows the overall trend over the years. 

In the absence of an absolute reference, the identification of noise sources at the precision limit can be awkward.  Tests for consistency were made by examining the stability of the supposedly stable pseudocontinua.  This identified a period of a few weeks in 2018 with enhanced fluctuations, of unknown origin, but simultaneously seen in different spectral regions, as described in Appendix A.  As a precautionary measure, all data from those weeks were removed from the further analysis.

\subsection{Selection of spectral lines}

Spectral line categories to be selected should be plausible candidates for identifying solar microvariability, probe different sections of the photosphere, and be subject to conditions of formation that likely can be modeled.  Chromosperic lines, in particular \ion{Ca}{ii}~H\,\&\,K, have of course been monitored since long ago, and in several other lines activity indices are seen that correlate with those.  \citet{thompsonetal20} used HARPS-N spectra to compare epochs of high and low solar magnetic activity, finding that the depths of some spectral lines significantly correlate with the \ion{Ca}{ii}~H\,\&\,K emission.   Related phenomena are seen for chromospherically more active K-type stars.  There, \citet{thompsonetal17} and \citet{wiseetal18} examined HARPS spectra to identify activity indicators that correlate with \ion{Ca}{ii}, extended by \citet{wiseetal22} by carrying out some stellar atmospheric modeling.  Identifying the spectral contributions from various solar surface features, \citet{cretignieretal24} concluded that the temporal variations of the \ion{Ca}{ii} index typically come about 70\% from plages, 26\% from network and 4\% from spots.  In an analogous vein, the main contributions to jittering in radial velocities are found to be the suppression of convective blueshift, not the presence of sunspots \citep{haywoodetal16, laklelandetal24, meunieretal10, senrajaguru23}.

Photospheric activity measures related to radial velocities were searched for by \citet{davisetal17}.  They used simulated disk-integrated time-series spectra to demonstrate that different absorption lines respond to activity in non-uniform ways, and therefore averaging over numerous different lines will necessarily wash out information. They also show that higher spectral resolution better retrieves information content from spectra that have been affected by stellar activity and enables better segregation of activity and planetary signals. 

The line selection should enable to examine primarily photospheric quantities, ideally such that would not strongly correlate with the classic parameter of \ion{Ca}{ii}~H\,\&\,K, and thus might carry independent information.  While chromospheric indices do outline the solar activity cycle and also reflect the overall changes of the average solar radial velocity from month to month, their detailed correlation with radial-velocity fluctuations breaks down on levels of m\,s$^{-1}$ (Fig.\ \ref{fig:solaractivity}), for which some other measures plausibly could exist to serve as proxies.  Below, the motivations for selection of lines from different species are described, together with discussions of what types of atmospheric phenomena that likely would be sampled by them.

\subsection{Selection of line parameters}

The choice of parameters to be measured is influenced by both the spectrometer performance, and by what could be amenable to later theoretical simulations.  For example, the modeling in Paper~I found that the jittering in radial velocity, as driven by fluctuations in granular convection, differs somewhat in amplitude between lines of different strength and in different wavelength regions.  However, the noise levels in present spectra do not permit to identify such differential radial-velocity signatures, especially not for limited subsets of selected lines.  Realistically measurable parameters, in which also tiny variations would likely be seen, include the basic line properties of strength, depth, and width.  Among these, the absorption equivalent width appears to be the most stable quantity for both observation and theory.  Any theoretically modeled line-depth would need to be adjusted for not only solar rotation but also for the finite spectrometer resolution (Fig.\ \ref{fig:520nm_general}), and fluctuations in width would need to be verified against instrumental profile stability.  The equivalent width is, at least to a first approximation, insensitive to minor variations in instrumental profile, even if its exact value depends on the level of spectrometer straylight and on how the continuum level is set. 
Likewise, the small [bi]annual modulation of the solar spectrum caused by the Earth's orbital motion should be negligible at the expected precision levels.  Of significance will be only the relative changes in equivalent width since absolute line strengths cannot theoretically be very precisely computed, being dependent on incompletely known laboratory oscillator strengths and the exact computational treatments of radiative transfer.

Foreseeing future hydrodynamic 3D modeling, the spectral features should be tightly defined in wavelength. In order to minimize noise from possible wavelength displacements in synthetic spectra, measurements of absorption equivalent widths are made between points in the spectrum with a minimal gradient, occasionally measuring over adjacent similar lines, and avoiding measurement boundaries in blends (Figs.\ \ref{fig:520nm_general} and \ref{fig:fe_line_groups}).  While such truncation would perhaps not be necessary for spectra from steadily calibrated radial-velocity spectrometers, it might be an issue for synthetic spectra, where spectral-line wavelength errors from laboratory measurements may be no longer negligible.  Line absorption is measured relative to the chosen pseudocontinuum segments which of course do not reach the exact continuum level, but for measurements of relative variability, that slight difference is of no concern.

No precisely stable continuum fitting is feasible over broader spectral regions spanning different {\'e}chelle orders, recorded during days of possibly variable and chromatic atmospheric extinction.  To obtain continuum levels, relative to which absorption lines are measured, multiple segments of pseudocontinua were selected.  As far as possible, these are chosen close to, and symmetric in wavelength about the lines to be measured (Figs.\ \ref{fig:520nm_general}, \ref{fig:fe_line_groups}, \ref{fig:chromospheric_lines}, \ref{fig:g-band_harps}, \ref{fig:mn_hyperfine_lines}, \ref{fig:balmer_lines}).  Compromises have to be made how clean these segments can be chosen, considering the need to span many pixels to limit photometric random noise, and also to be close to the target line to limit possible systematic drifts.

\section{Photospheric \ion{Fe}{i} and \ion{Fe}{ii} lines} 

Fe lines are ubiquitous throughout the spectrum and are commonly used for various atmospheric diagnostics, velocity shifts, and other.  As listed in Table \ref{table:linelist1}, \ion{Fe}{i} and \ion{Fe}{ii} line features were selected from various parts of the spectrum, sometimes spanning over multiple lines (Fig.~\ref{fig:520nm_general}).  The line selection is essentially the same as used in the theoretical modeling of Paper~I, however, with slight differences caused by the lower spectral resolution here, as compared to the synthetic hyper-high one.  The lines are grouped as \ion{Fe}{i} 430-445, \ion{Fe}{i} 520-535, \ion{Fe}{i} 670-685, and \ion{Fe}{ii} 435-475 nm.  

These four line groups represent different classes of Fe lines, as was also seen from their dissimilar radial-velocity behavior in Paper~I.  The drifts in their absorption strength are shown in Fig.~\ref{fig:all_fe_lines_abs}, plotted as a function of the chromospheric \ion{Ca}{ii} activity index and the radial velocity, parameters that also well segregate the changes from year to year.  
Here, as well in later figures, the values in each line group are normalized to 100\% of the arithmetic average for their full 3-year dataset. 

Similar to the radial-velocity jittering in Paper~I, the greatest amplitudes are seen for the \ion{Fe}{ii} lines at short wavelengths, decreasing for \ion{Fe}{i} and at longer wavelengths.  With increasing activity, the line absorption becomes weaker (although the noisier data for the reddest lines are somewhat inconclusive).  When plotted against radial velocity, the patterns remain basically similar, although now with a somewhat greater dispersion.

\section{The green \ion{Mg}{i} triplet and \ion{Mg}{i} $\lambda$\,457.1 nm}

A group of strong lines with formation in the higher photosphere (with contributions from also the lower chromosphere) is the green magnesium triplet of \ion{Mg}{i} b{$_1$} $\lambda$\,518.3, \ion{Mg}{i} b{$_2$} $\lambda$\,517.2, and \ion{Mg}{i} b{$_3$} $\lambda$\,516.7 nm.   Their atomic energy levels are coupled such that they are transitions between one common upper level and three different lower levels with the quantum numbers J=2, J=1, and J=0 for  b{$_1$}, b{$_2$}, and b{$_3$}, respectively.  Their Grotrian term energy diagram is shown in, e.g., \citet{alexeevaetal18} and \citet{peraltaetal22}. Another tantalizing target is the intercombination line \ion{Mg}{i} $\lambda$\,457.1 nm.  Its upper energy level is the same as the lower level of the \ion{Mg}{i} b{$_2$} line, from where it transits to the ground level.  The selected wavelength intervals are in Table \ref{table:linelist2}.  

\subsection{The \ion{Mg}{i} b triplet}

The formation of the \ion{Mg}{i} b-line triplet in the Sun and in other stars has been discussed by multiple authors.  Their line wings probe the upper photosphere while the line cores form in the lower chromosphere.  Their shared upper energy level contributes to the complexity of their formation.  Since both source functions and opacities are affected by deviations from local thermodynamic equilibrium, LTE (mainly through photoionization), these lines are not particularly sensitive to the local atmospheric structure \citep{sassoetal17, zhaoetal98}.  However, the triplet can be seen as a diagnostic for stellar activity \citep{sassoetal17} and was studied for photospheric changes in G- and K-type stars by \citet{basrietal89}.  Observed and simulated solar surface images in \ion{Mg}{i} b{$_2$}, and its line formation within photospheric granulation is evaluated by \citet{ruttenetal11}.   Various aspects of \ion{Mg}{i} lines in the Sun and cool stars, including effects of departures from LTE, are further discussed by \citet{alexeevaetal18, carlssonetal92, osoriobarklem16, peraltaetal22, peraltaetal23, sassoetal17, zhaoetal98}, and others.
 
\subsection{The intercombination line \ion{Mg}{i} 457.1 nm}

This rather special, so-called “semi-forbidden”, intersystem line originates between the atomic levels 3s{$^2$} ${^1}$S${_0}$ – 3p {$^3$}P${_1}$.  As compared to the \ion{Mg}{i} b triplet, its formation is expected to be relatively simple.  Already very early non-LTE calculations showed it to be completely dominated by thermal excitation \citep[e.g.,][]{altrockcanfield74, altrockcannon72, mauasetal88}. Due to the dominance of collisional processes in forbidden lines, its source function is tightly coupled to the local temperature.  In one-dimensional model atmospheres, the line forms at around 500 km height, and thus provides temperature diagnostics from around the temperature minimum, although some non-LTE effects could still be present \citep{langangencarlsson09, sassoetal17}.  At solar disk center, its lack of a central reversal constrains the lowest possible position of the chromospheric temperature rise, while \citet{carlssonetal92} point out that it is the only line in the optical spectrum producing a line center emission reversal at the solar limb (where the line-of sight crosses the temperature minimum).  Despite its modest atomic oscillator strength, the absorption line is rather strong because its opacity is determined by the relatively high population of its lower energy level -- the \ion{Mg}{i} ground state. 

\subsection{Relations among \ion{Mg}{i}  lines}

\begin{figure}
\centering
 \includegraphics[width=\hsize]{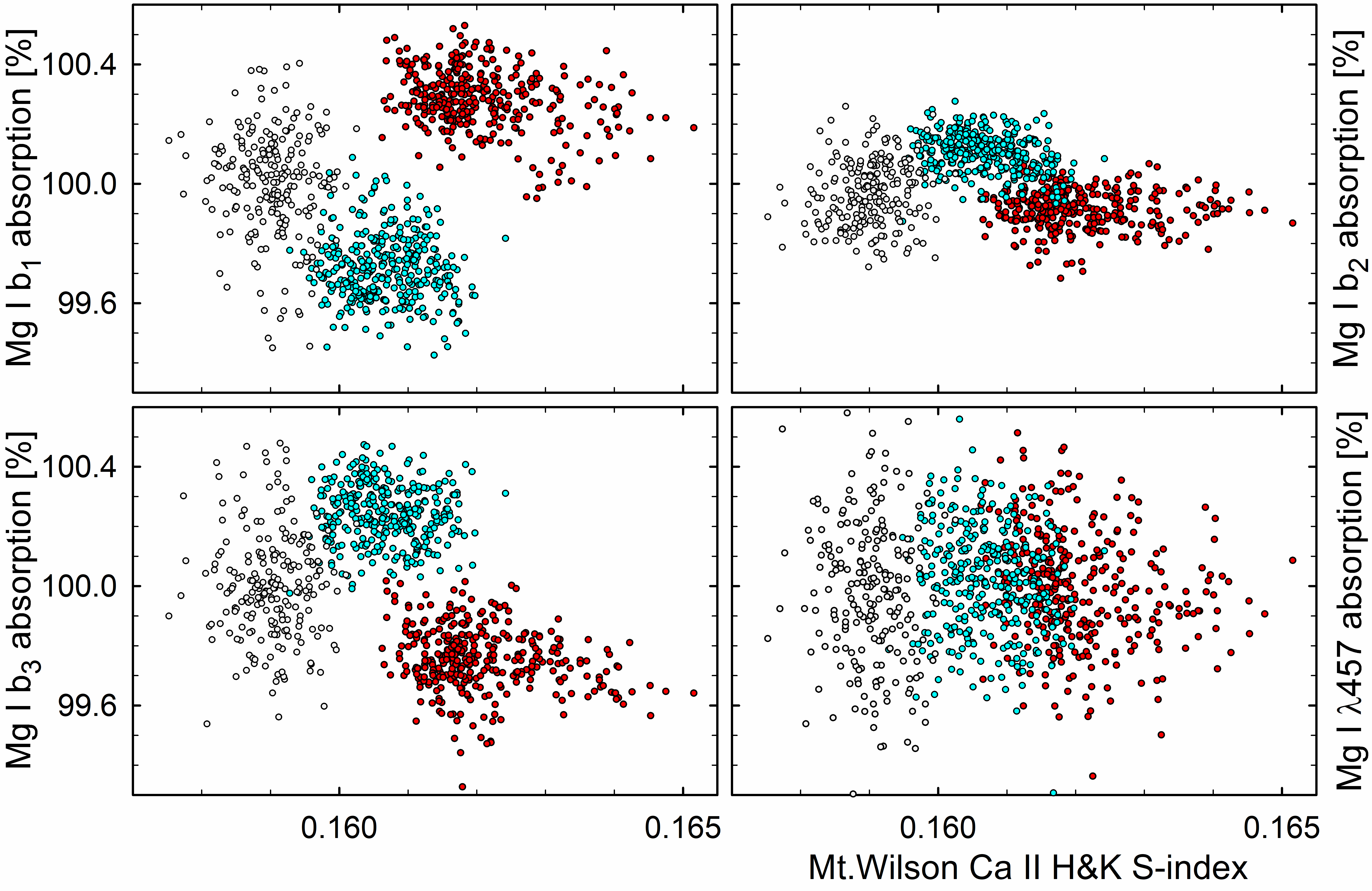}
     \caption{Relative absorption equivalent width changes in \ion{Mg}{i} lines during the selected seasons of 2016-2017-2018 (red-cyan-gray), as function of the Mt.Wilson \ion{Ca}{ii} H\,\&\,K S-index.  }   
\label{fig:mg_b1-b2-b3_abs-mtw}
\end{figure}

\begin{figure}
\centering
 \includegraphics[width=\hsize]{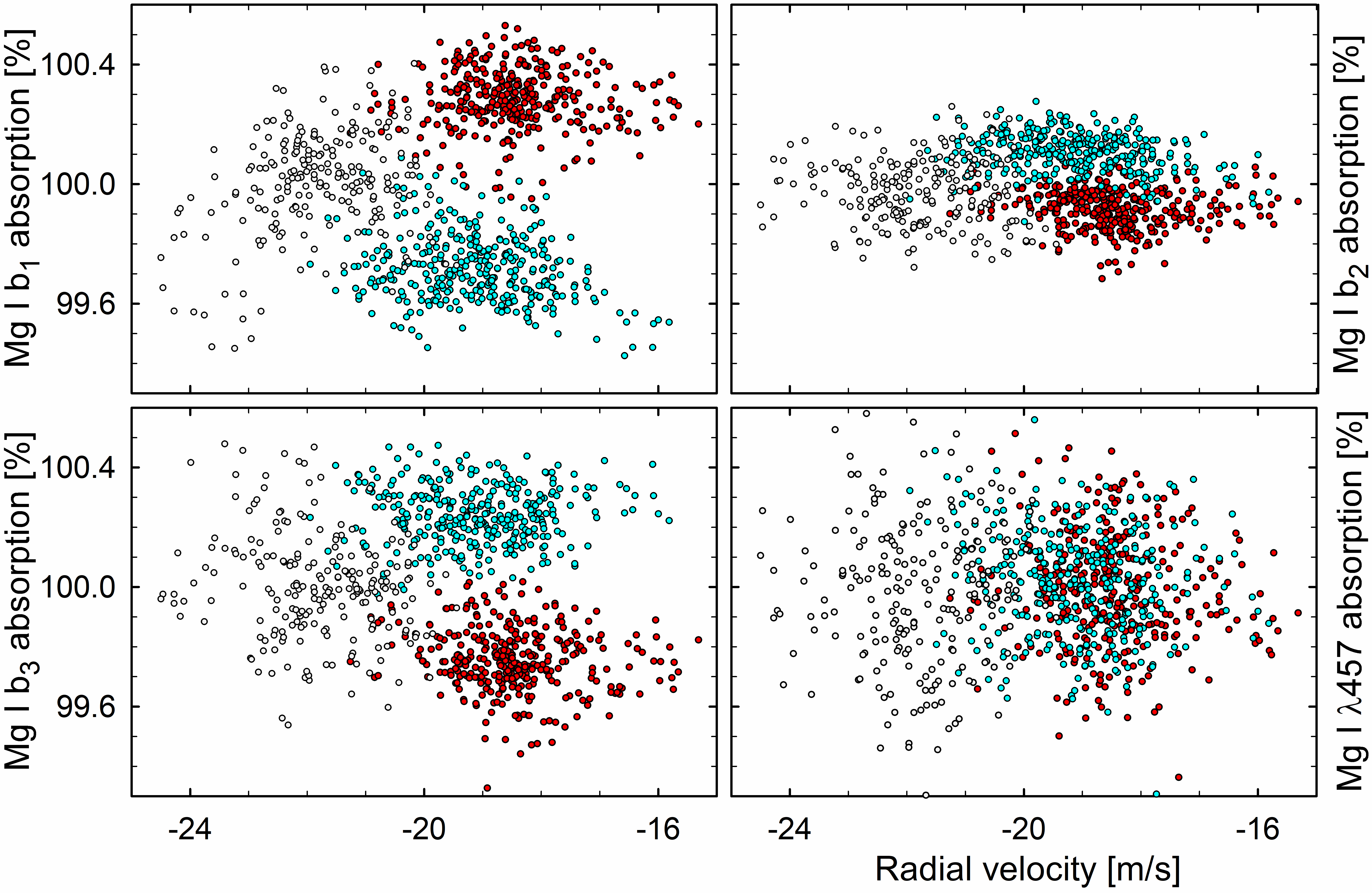}
     \caption{Relative absorption equivalent width changes in \ion{Mg}{i} lines during the selected observing seasons of 2016-2017-2018 (red-cyan-gray), as function of the radial velocity for the full spectrum.  }   
\label{fig:mg_b1-b2-b3-457_abs-radvel}
\end{figure}

\begin{figure}
 \includegraphics[width=\hsize]{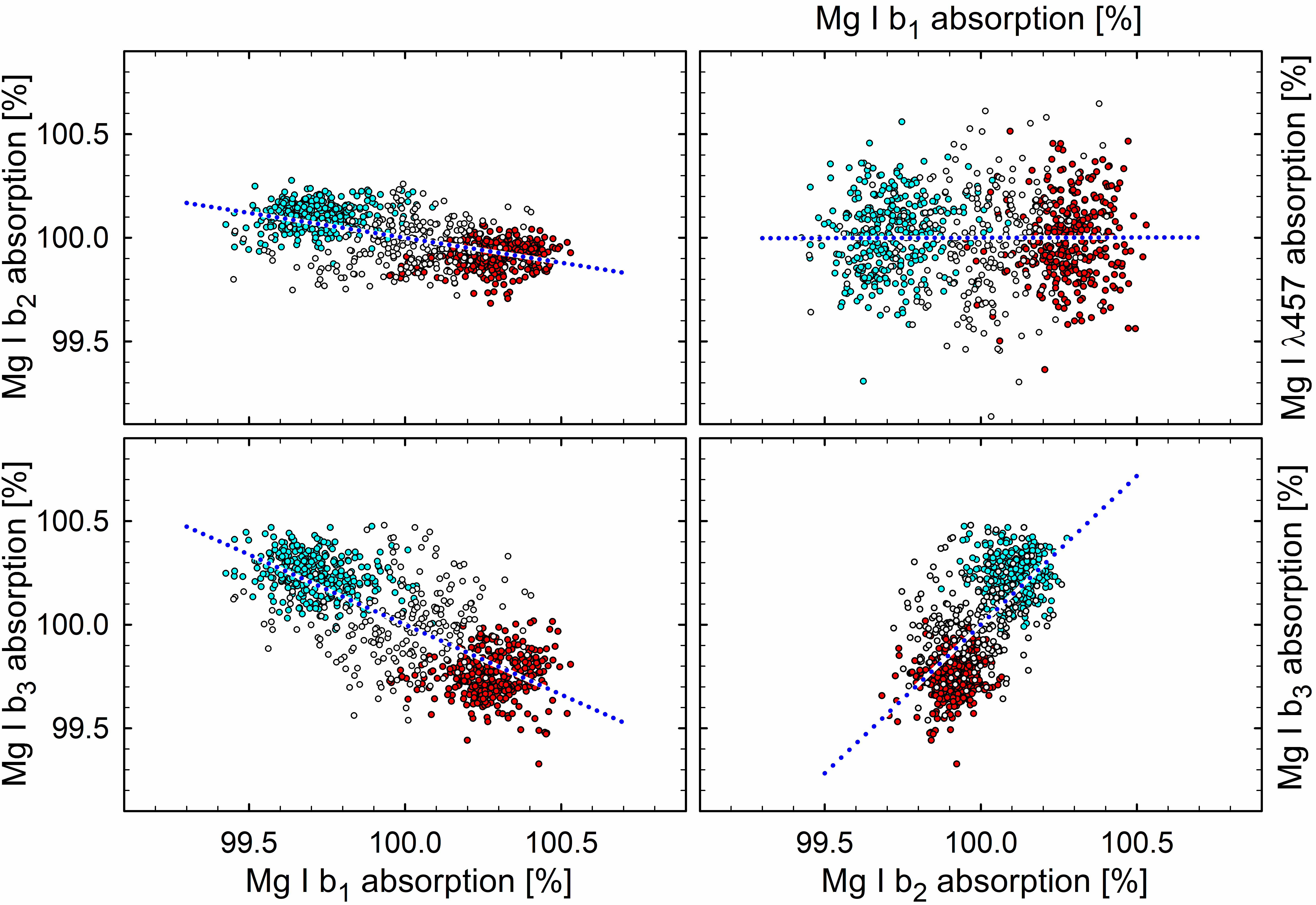}
     \caption{Relative changes in the absorption equivalent widths of the various \ion{Mg}{i} lines during the selected seasons of 2016-2017-2018 (red-cyan-gray).  Dotted blue lines are fits to the data.  }   
\label{fig:mg_b1-b2-b3-457}
\end{figure}

Figure \ref{fig:mg_b1-b2-b3_abs-mtw} shows the dependences of \ion{Mg}{i} line absorptions as function of the \ion{Ca}{ii} chromospheric index, and Fig.\ \ref{fig:mg_b1-b2-b3-457_abs-radvel} as function of the radial velocity.   The trends among the three \ion{Mg}{i} b lines are distinctly different and neither do they follow the \ion{Ca}{ii} index.  (However, their behavior reminds of those for three Balmer lines to be discussed later.)  The line strengths vary between the observing seasons, with an anticorrelation between the strongest b{$_1$} line, and the two others.  The relative changes among the lines show clear and systematic patterns (Fig.~\ref{fig:mg_b1-b2-b3-457}), with the steepest mutual dependence between b{$_2$} and b{$_3$}.  Although it perhaps appears striking on the plots, the total amplitudes amount to only about half a percent in absorption equivalent widths. 

The data for the weaker \ion{Mg}{i} $\lambda$\,457.1 nm line are noisier (reflecting its narrower width; Fig.\ \ref{fig:chromospheric_lines}) and show no change between the years (Figs.~\ref{fig:mg_b1-b2-b3_abs-mtw}-\ref{fig:mg_b1-b2-b3-457_abs-radvel}).  Neither does this special line show any correlation with the contemporaneous \ion{Mg}{i} b strengths (Fig\ \ref{fig:mg_b1-b2-b3-457}), and this semi-forbidden line thus appears to be in a class by itself.

\begin{figure*}
\sidecaption 
 \includegraphics[width=12.9cm]{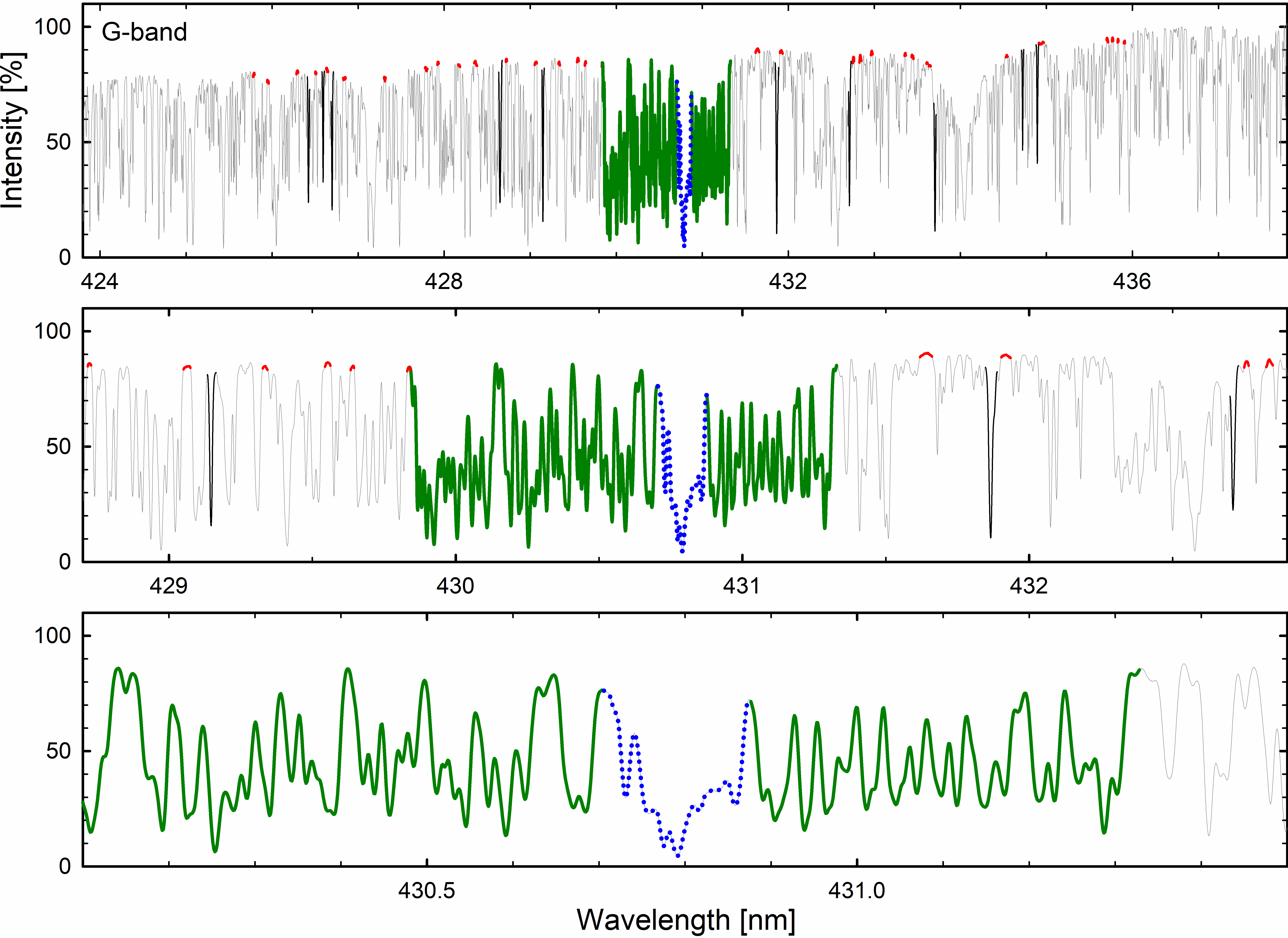}
     \caption{The {\it{G}}-band region together with selected spectral features as observed with HARPS-N: {\it{G}}-band core (dotted blue), full {\it{G}}-band (dotted blue + green), surrounding nearby \ion{Fe}{i} lines (black), and pseudocontinuum reference points (red). An averaged 100-exposure spectrum is plotted in thin gray.  From top down, successively narrower spectral segments are shown. }  
\label{fig:g-band_harps}
\end{figure*}

\begin{figure}
\centering
 \includegraphics[width=\hsize]{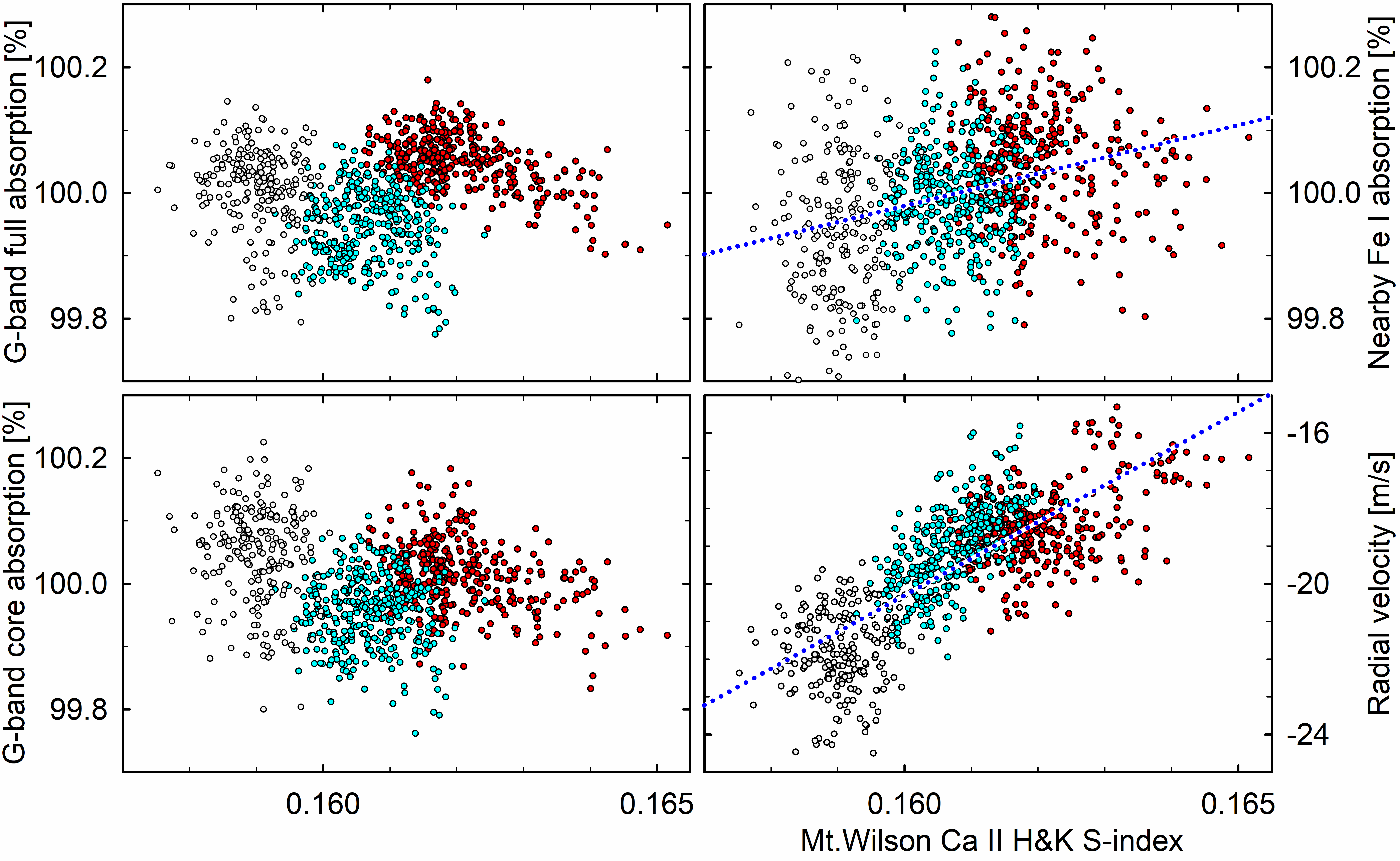}
     \caption{Absorption equivalent widths for the full {\it{G}}-band region, its central core, and nearby \ion{Fe}{i} lines, as function of the \ion{Ca}{ii}~H\,\&\,K S-index. The different seasons of 2016, 2017, and 2018 are marked in red, cyan, and gray.   Also, the corresponding radial velocity variations for the full spectrum are shown.  For cases with apparent systematic dependences, dotted blue lines show fitted relations. }  
\label{fig:g-band_abs_mtw}
\end{figure}

\section{Magnetically influenced features}

\subsection{The {\it{G}}-band}

The {\it{G}}-band is a heavily blended region around 430 nm, with many rotational and vibrational transitions from the CH molecular bandhead, intermingled with numerous atomic lines.  Its spectral appearance was described in Paper~I in connection with its modeling in non-magnetic and magnetic granulation.  

Much of the small-scale magnetic fields outside sunspots is outlined by the bright network, which at high spatial resolution resolves into solar filigree (sometimes called bright points, although not really point-like), occupying part of the spaces between granules. The corresponding appearance in the chromosphere is more smeared out, apparently reflecting the expansion of magnetic flux into higher layers.  Given that convective blueshift is suppressed in magnetically disturbed granulation, and because of their large areal extent (also far outside sunspot groups), it has been realized that the changing area coverage of the magnetic network and its associated plages is responsible for a great fraction of the solar-cycle modulation in apparent radial velocity, quantitatively confirmed by \citet{laklelandetal24} and \citet{meunieretal10}.  

The magnetic network of the quiet Sun appears bright in various spectral regions, in particular in molecular and other temperature-sensitive lines.  Among these, observations and modeling in the {\it{G}}-band are particularly extensive (Paper~I).  Since the {\it{G}}-band emission thus traces the network, it appears plausible that the {\it{G}}-band brightness could be a proxy for the magnetic network in integrated sunlight.   

Figure \ref{fig:g-band_harps} shows the {\it{G}}-band spectrum, as measured by HARPS-N.  The adopted boundaries are the same as in the synthetic spectra at hyper-high resolution in Paper~I, with the full {\it{G}}-band spanning 429.84--431.33 nm, and its central core 430.71--430.88 nm (Table \ref{table:linelist2}).  Also, the ten nearby \ion{Fe}{i} lines (Table \ref{table:linelist1}) and the 32 pseudocontinuum reference points are identical.  A comparison with Fig.\,12 in Paper~I demonstrates effects of different spectral resolutions.

The absorption in these bands is shown in Fig.~\ref{fig:g-band_abs_mtw}.  Over the years, the systematic changes in both the full {\it{G}}-band, and in its core are quite similar to those in \ion{Mg}{i} b{$_1$}, the strongest line of the triplet, but differ from those in the weaker \ion{Mg}{i} b lines (Fig.~\ref{fig:mg_b1-b2-b3_abs-mtw}).  The nearby \ion{Fe}{i} lines that crowd around the {\it{G}}-band strengthen with increased magnetic activity, a trend opposite to that for the more isolated photospheric \ion{Fe}{i} lines seen in Fig.~\ref{fig:all_fe_lines_abs}. 

\begin{figure}
\centering
 \includegraphics[width=\hsize]{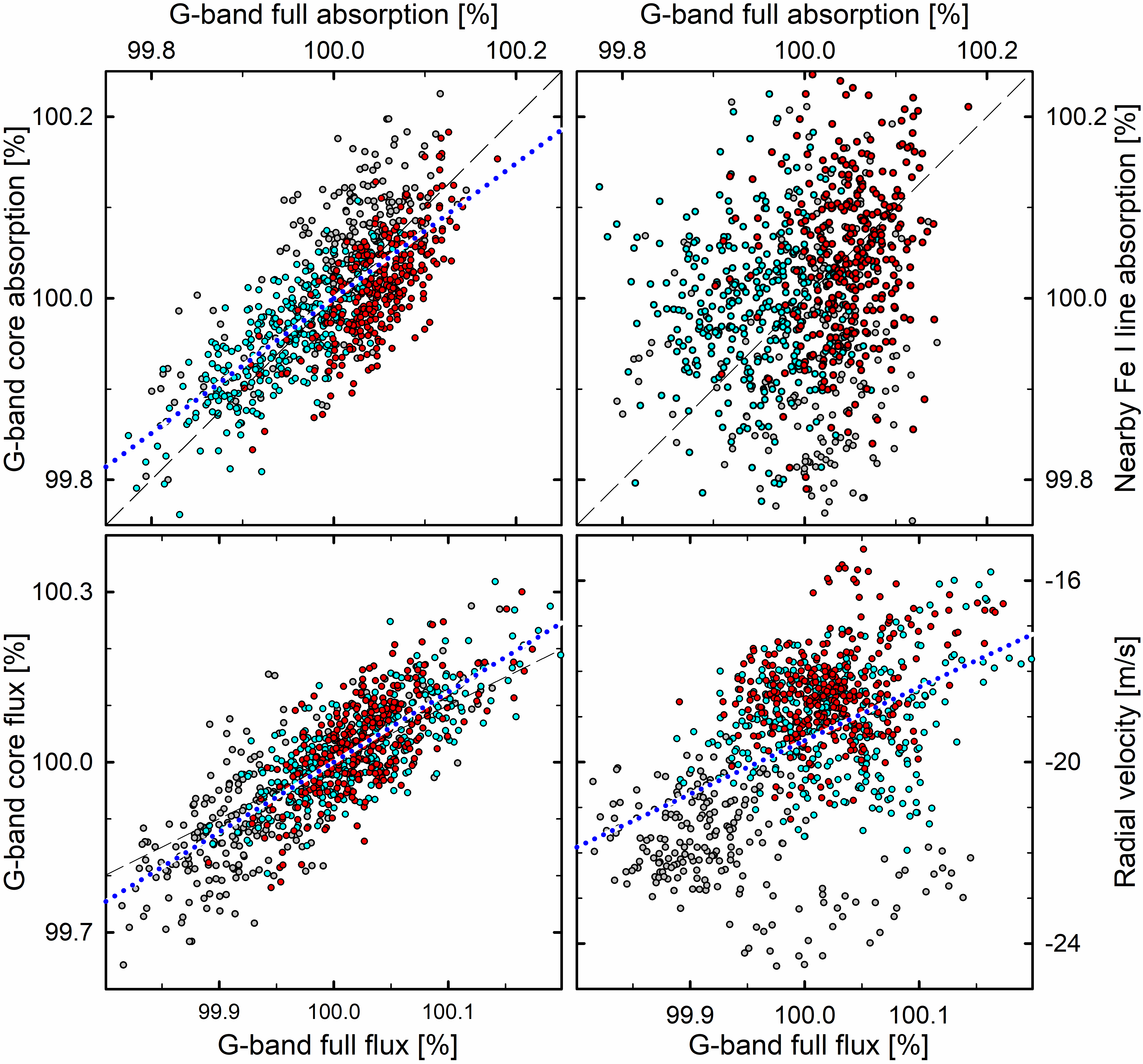}
     \caption{Ratios between the full {\it{G}}-band fluxes and its absorption equivalent widths, the {\it{G}}-band core, and that of nearby \ion{Fe}{i} lines.  Different seasons of 2016, 2017 and 2018 are marked in red, cyan, and gray.  Fitted relations are dotted blue lines, while the identity relations are dashed black. The core flux shows a steeper relative variability than the full band.   }  
\label{fig:g-band_relations}
\end{figure}

The flux in the {\it{G}}-band core is more variable than that of the full {\it{G}}-band.  As examined in Paper~I, the core has enhanced sensitivity to both spatial and magnetic variations, and such a variability difference is thus expected.  Figure \ref{fig:g-band_relations} shows the measured {\it{G}}-band quantities both as absorption and as flux, the latter corresponding to what is observed in monochromatic solar images.  Similar to the dependence for \ion{Ca}{ii}~H\,\&\,K, there is a correlation with simultaneous variations in radial velocity, although the relation shows more scatter (Figs.~\ref{fig:g-band_abs_mtw}--\ref{fig:g-band_relations}).

Enhanced contrast in magnetic fluxtubes is produced also in other spectral features, in particular molecular lines. As remarked in Paper~I, the violet CN band around 388 nm gives a contrast about 1.4 times greater than the {\it{G}}-band, although observationally that short-wavelength region is more demanding (and also falls shortward of the HARPS-N spectral range).  Signatures from CN should represent the low chromosphere.  A narrow feature of the CN 388.3 nm bandhead was included in the Kitt Peak monitoring program, showing a full solar-cycle peak-to-peak amplitude of $\sim$~3\% \citep{livingstonetal07}, an order of magnitude smaller than the corresponding change in the \ion{Ca}{ii}~H\,\&\,K index.

\begin{figure*}
\sidecaption
 \includegraphics[width=12.9cm]{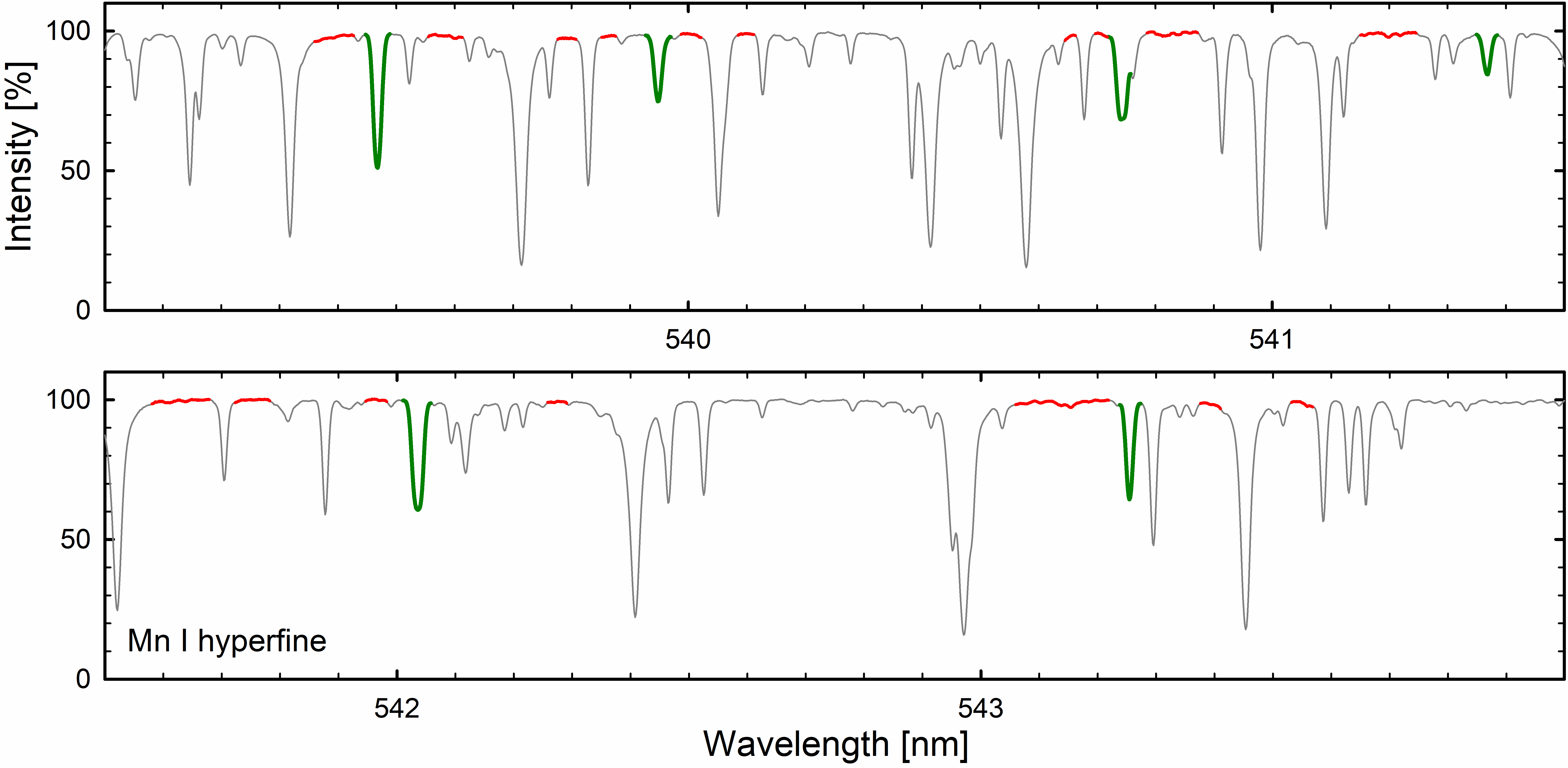 }
     \caption{Examples of selected \ion{Mn}{i} lines with hyperfine splitting in the 540 nm region with pseudocontinuum segments marked in red.  The lines are noticeably broader than others of comparable depth. The full set of sampled \ion{Mn}{i} lines extends over the range 511 to 644 nm (Table~\ref{table:linelist2}). }  
\label{fig:mn_hyperfine_lines}
\end{figure*}

\begin{figure}
\centering
 \includegraphics[width=\hsize]{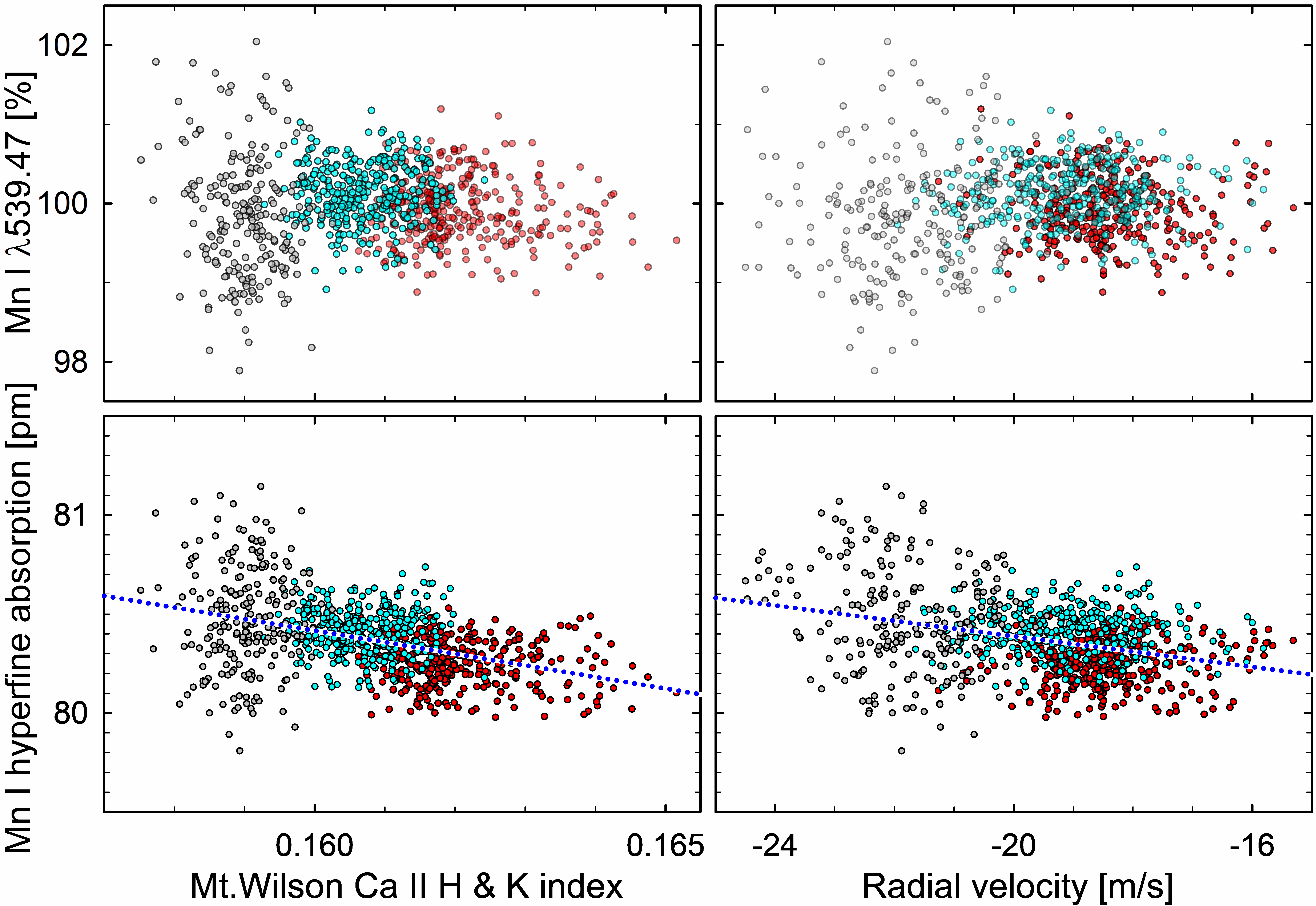}
     \caption{Variations in hyperfine split \ion{Mn}{i} lines.  Data for the previously studied single line $\lambda$\,539.47 nm (top row) are inconclusive but the summed equivalent widths of 16 different \ion{Mn}{i} lines reveal a clear trend (bottom).  Data for 2016-2017-2018 are shown in red-cyan-gray  }  
\label{fig:mn_hyperfine_mtw_radvel}
\end{figure}

\begin{figure*}
\sidecaption
 \includegraphics[width=12.9cm]{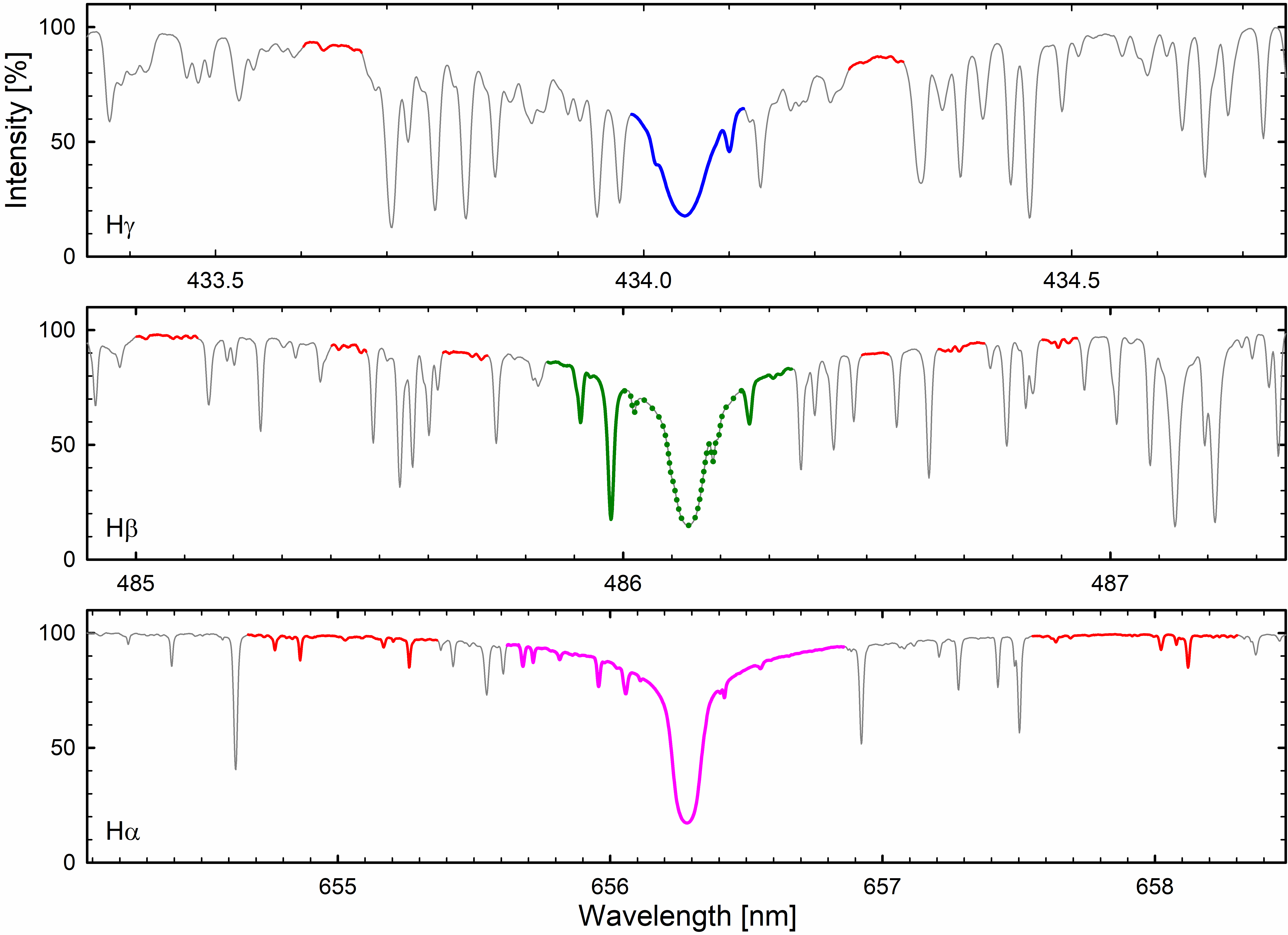}
     \caption{Spectral regions around H$\alpha$, H$\beta$, and H$\gamma$.  Solid bold lines denote the line intervals measured; for H$\beta$ its central core was also measured separately (dotted).  In each case, lines are measured relative to local pseudocontinua marked in red, chosen symmetrically about the line centers.  The average spectrum from 100 HARPS-N exposures is plotted as a thin dark line.  The intensities are normalized to each local pseudocontinuum. }  
\label{fig:balmer_lines}
\end{figure*}

\subsection{Hyperfine split \ion{Mn}{i} $\lambda$\,539.47 nm} 

Another class of lines that may carry particular signatures from photospheric magnetic regions are those whose atomic structure causes hyperfine splitting.  This causes broad, intrinsically wide and flat-bottomed profiles, for which the line broadening is quite insensitive to nonthermal motions such as the wavelength smearing by granular velocities.  This makes them suitable as temperature indicators, a property examined already by \citet{elsteteske78}, \citet{elste86}, \citet{erkapicvince93}, and \citet{vincevince03}.  

A prominent line is \ion{Mn}{i} $\lambda$\,539.47 nm, with a hyperfine broadening comparable to its thermal one.  At Kitt Peak, its measurement in integrated sunlight began in 1979.  It turned out to vary differently from other features, showing the greatest solar-cycle variation \citep{livingstonetal07, livingstonwallace87}.  Its equivalent width was measured to change from about 7.86 pm at solar activity maximum to about 7.98 pm at minimum, an amplitude of $\sim$1.5\%.  This line was later included in a full-disk monitoring program in Belgrade \citep{arsenijevicetal88, vinceetal88}, confirming the solar-cycle amplitude \citep{danilovicvince04, danilovicetal05, skuljanetal92, skuljanetal93}.  

Suggested explanations included optical pumping from the ultraviolet \ion{Mg}{ii}~k $\lambda$\,279.55 nm that happens to almost coincide with a transition in Mn \citep{doyleetal01}, but later analyses by \citet{danilovicetal16} and \citet{vitasvince05, vitasvince07} showed this not to be significant.  However, the temperature sensitivity of this \ion{Mn}{i} line causes its absorption to weaken less in normal granulation and monochromatic images therefore show a bright network, similar to an unsigned magnetogram \citep{malanushenkoetal04, vinceetal05}. That its line parameters stand out in contrast to nearby \ion{Fe}{i} lines is seen also in solar spectra from the center to the limb \citep{osipovvasilyeva19} while the solar-cycle variations can be modeled with changing distributions of solar magnetic features \citep{danilovicvince05, danilovicetal16}.  As opposed to other lines often used in solar magnetometry, \ion{Mn}{i} $\lambda$\,539.47 nm weakens with increasing field strength, making it sensitive to also weaker fields in the quiet Sun \citep{lopezaristeetal02, sanchezalmeidaetal08, vitasetal09}. 

Its behavior was further clarified in 3D modeling and spectral synthesis by \citet{vitasetal09}.  The line weakens in intergranular magnetic concentrations while the corresponding effect in ordinary and narrower photospheric lines has less impact in full-disk flux due to their decreased line depth in normal granulation.  Somewhat analogous effects exist for the {\it{G}}-band, and in the extended wings of strong lines such as H{$\alpha$} or H{$\beta$} \citep{leenaartsetal06a, leenaartsetal06b, leenaartsetal12}.  For other late-type stars, 3D non-LTE calculations for \ion{Mn}{i} are by \citet{bergemannetal19}.  As compared to Fe, Mn is particularly sensitive to non-LTE effects because of a low cosmic abundance, great photo-ionisation cross-sections, and a somewhat special atomic energy level structure.

The extent to which the \ion{Mn}{i} $\lambda$\,539.47 nm could be affected by telluric absorption was examined by \citet{vincevince10}.  Its core was found to be free from such contamination although tellurics might affect unsuitably chosen local continuum regions.

\subsection{Other hyperfine split \ion{Mn}{i} lines} 

With such behavior, the detailed variability in lines such as \ion{Mn}{i} $\lambda$\,539.47 nm might well carry specific signatures about photospheric magnetic conditions connected to also radial-velocity fluctuations.  In the case of our HARPS-N data, an issue arises as to how small fluctuations reliably can be measured.  If guided by previous Kitt Peak and Belgrade measurements, the change over our current fraction of a solar activity cycle can be expected to be only a fraction of one percent.  Not being especially strong  (Fig.~\ref{fig:mn_hyperfine_lines}), the line extends over only a limited wavelength range, and the noise level makes it hard to deduce a likely physical signal, as seen on the top row of Fig.~\ref{fig:mn_hyperfine_mtw_radvel}. 

Some other lines were considered as candidates for measurement but, although also a few other atomic species display hyperfine structure, those from \ion{Mn}{i} are the most distinct.  \citet{lopezaristeetal02} list \ion{Mn}{i} lines with conspicuous broadenings related to their hyperfine structure.  Of their 19 lines, 16 fall within the HARPS-N spectral range, and were chosen to represent a more general hyperfine splitting signature (Table \ref{table:linelist2}).  Their equivalent widths were measured individually relative to pseudocontinua around each of them, and their summed equivalent width of about 80 pm (800 m{\AA}) is shown on the bottom row of Fig.~\ref{fig:mn_hyperfine_mtw_radvel}.

The signal summed from these 16 lines now shows a clear trend during 2016-2017-2018 as an increase of the absorption by {$\sim$}0.3~\% when going toward years of lower solar activity, consistent with the trends seen earlier at Kitt Peak and Belgrade for their single line.  However, perhaps contrary to expectations, the trends are qualitatively similar and actually smaller compared to what was seen for more ordinary \ion{Fe}{i} and \ion{Fe}{ii} lines in Fig.~\ref{fig:all_fe_lines_abs}.

\subsection{Unsigned magnetic flux and line formation}

The variability of several spectral lines is influenced by the (unsigned) magnetic flux \citep{rutten19}.  That can be measured from (unpolarized) intensity spectra by fitting the line-broadening dependence among lines with known Land{\'e} $\varg$$_{\textrm{eff}}$-factors for magnetic sensitivity.  For HARPS-N solar spectra this was done by \citet{lienhard23}, averaging over more than 4,000 lines.  With sufficient signal-to-noise, one would be able to examine lines with different Zeeman sensitivity and formed in different (and also differently moving) photospheric structures \citep[e.g.,][]{liuetal23}.  Line pairs with different Zeeman sensitivities but otherwise similar properties and formed under the same atmospheric conditions could provide more precise determinations \citep{smithasolanki17}.  

Magnetic fields in the quiet Sun are reviewed by \citet{bellotrubioorozcosuarez19} and \citet{sanchezalmeidamartinezgonzalez11}.  Different spectral lines have different sensitivities to differently strong magnetic fields \citep{quinteronodaetal21}, while near-infrared lines might be especially valuable for mapping fields into the deeper photosphere \citep{hahlinetal23, laggetal16}.  There thus exists a wealth of information about spectral-line properties in solar magnetic structures, however, not much of this can yet be utilized for the Sun-as-a-star because of constraining signal-to-noise and finite spectral resolution in even the best current radial-velocity instruments.

\begin{figure}
\centering
 \includegraphics[width=\hsize]{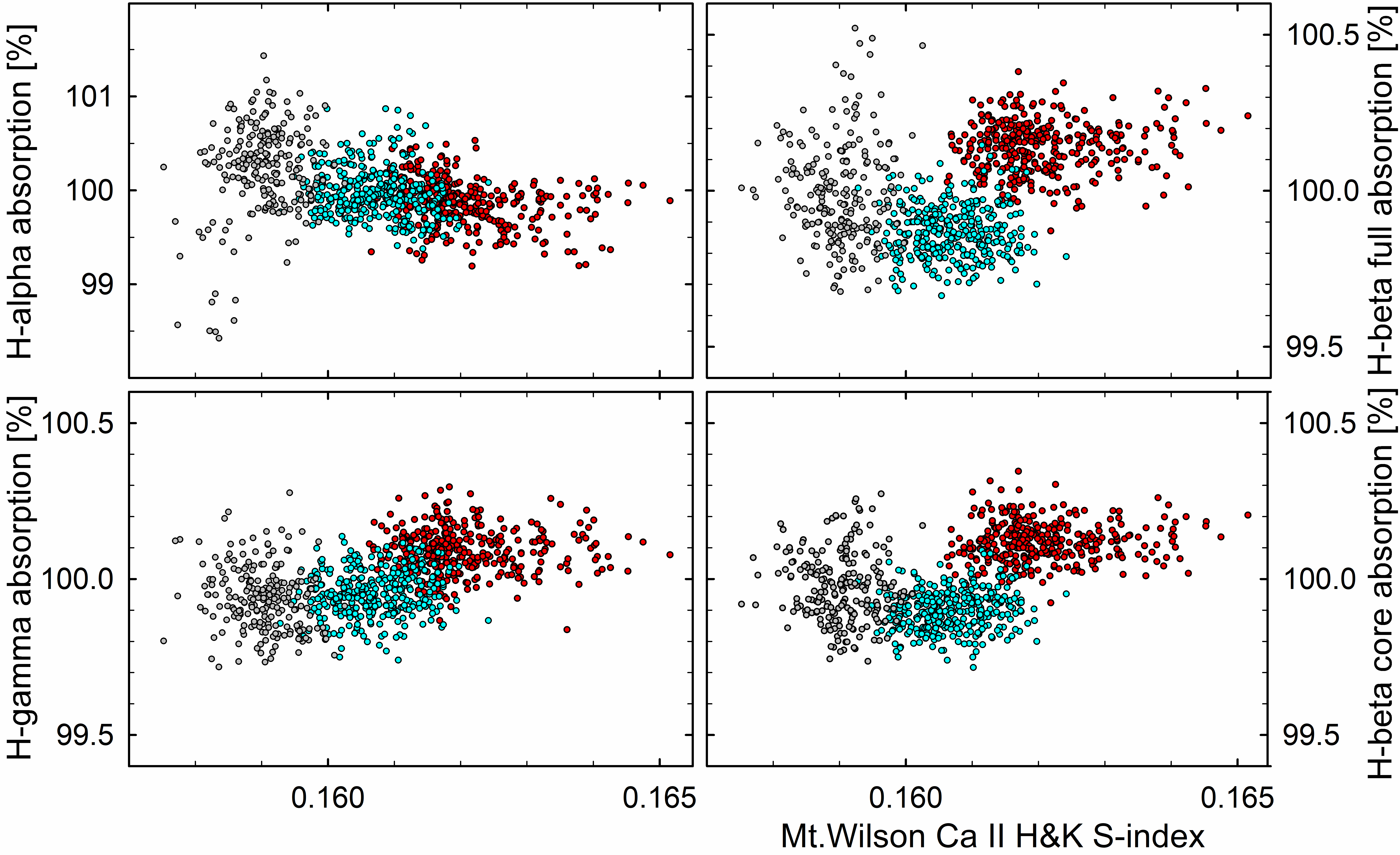}
     \caption{Relative absorption equivalent widths for the Balmer lines H$\gamma$, H$\beta$ (full and core), and H$\alpha$, as function of the Mt.Wilson \ion{Ca}{ii}~H\,\&\,K S-index.   Different colors indicate data from the observing seasons of 2016 (red), 2017 (cyan) and 2018 (gray).  Because of its greater variability, the vertical scale for H$\alpha$ is more extended. }  
\label{fig:ha_hb_hg_abs_mtw}
\end{figure}

\begin{figure*}
\sidecaption
 \includegraphics[width=12.9cm]{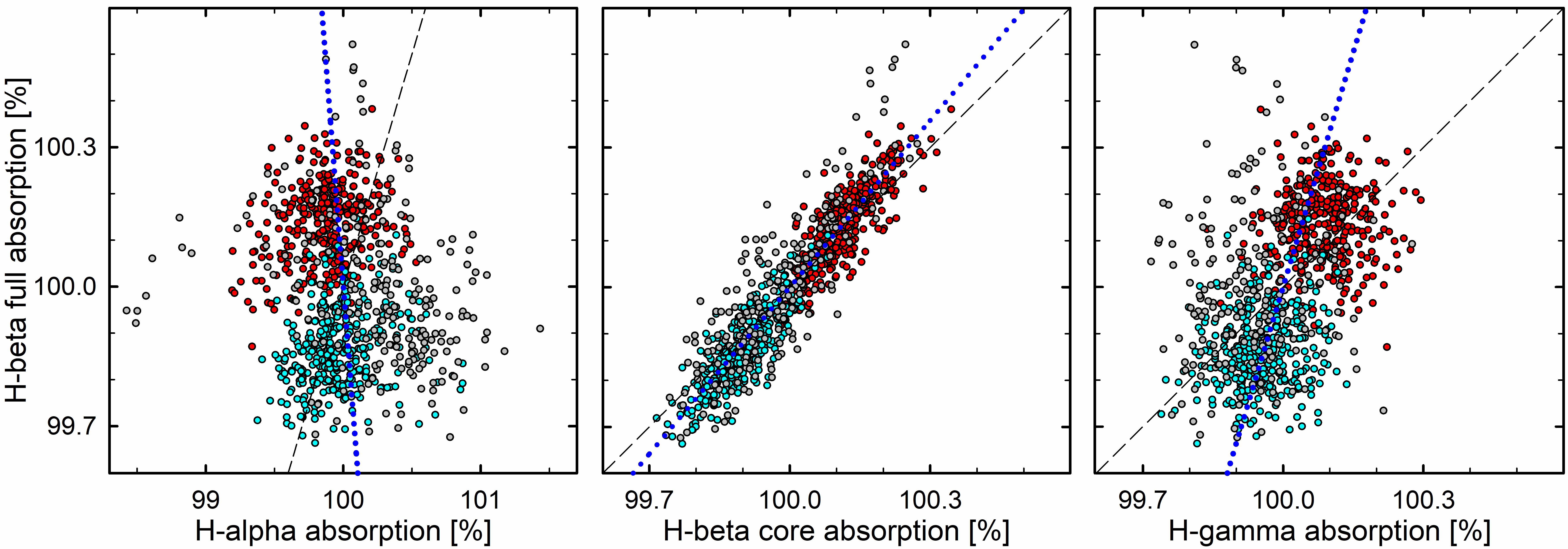}
     \caption{Ratios of the absorption equivalent widths for the Balmer lines H$\alpha$, the H$\beta$ core, and H$\gamma$, versus that of the full H$\beta$ (Fig.~\ref{fig:balmer_lines}).  Red, cyan, and gray indicate data from the observing seasons of 2016, 2017 and 2018. Dashed lines show the identity ratios while those in dotted blue are fits to the data.  Because of its greater variability, the horizontal scale for H$\alpha$ is more extended.  }  
\label{fig:balmer_ratios}
\end{figure*}

\section{The strongest lines}

Although we focus on photospheric features, also some stronger and more chromospheric lines are considered.  They may not provide precise radial velocities, but their measures connect with previous work for \ion{Ca}{ii}~H\,\&\,K, they display greater variability than purely photospheric lines, and they are useful as indicators for magnetic activity.  

We generally reference various measures to the Mt.Wilson S-index for \ion{Ca}{ii}~H\,\&\,K (Fig.\ \ref{fig:solaractivity}), obtained as the flux in the central (chromospheric) emission core, relative to that in the (mainly photospheric) flanks of the absorption lines. This quantity was examined by \citet{egelandetal17} while the definition \citep{duncanetal91} goes back to the specifics of the Mt.Wilson coudé spectrograph \citep{vaughanetal78}, later calibrated on also other spectrometers \citep{hartmannetal84, oranje83}. The \ion{Ca}{ii} emission has been studied by many; its solar-cycle variation by \citet{whitelivingston81} and line profiles in solar plages by \citet{oranje83}.  \citet{dumusqueetal21} describes it for HARPS-N spectra, and the values in our plots are from its data release.  Line components in solar surface structures were examined by \citet{cretignieretal24} and long-term variations by \citet{singhetal23}. On the smallest scales, bright grains appear to be produced by shocks in the mid-chromosphere \citep{carlssonstein97, vankootencranmer24}.

For the appearance of \ion{Ca}{ii}~H\,\&\,K in stars very similar to the Sun, see \citet{pasquinietal88} for $\alpha$~Cen~A (G2~V) and \citet{dravinsetal93} for $\beta$~Hyi (G2~IV).  A broader review of stellar chromospheric variability is by \citet{degrijskamath21}. 
 
\subsection{Balmer lines: H\,$\alpha$, H\,$\beta$, H\,$\gamma$}

Also H$\alpha$ can be used to indicate stellar activity.  Over solar active regions, its absorption core is partially filled in, and its flux is thus affected by chromospheric activity.  In G- and K-dwarfs and subgiants that was considered by \citet{cayreletal83, cincunequietal07, gomesdasilvaetal14, gomesdasilvaetal22, martinezarnaizetal10, pasquinipallavicini91, zarro83, zarrorodgers83}, and others.  The approach is more practical for active stars, where the filling-in is greater than the sub-percent values for integrated sunlight.  From an extensive sample of F-, G-, and K-stars observed with HARPS, \citet{gomesdasilvaetal22} studied how the signals depend on the chosen bandwidths for the flux, and how that varies with stellar activity level and metallicity.  Bandpass choices with a narrow central interval referenced to surrounding broader regions are mainly oriented toward an expected narrow central emissions in more active stars \citep{boisseetal09, bonfilsetal07, kursteretal03}.

The H$\alpha$ transition has its lower energy level at 10.2 eV, which means that the line opacity is very temperature dependent.  Because of hydrogen's low atomic weight, its thermal broadening is very large and the line is not a good diagnostic of nonthermal motions. The line width is instead a measure of the gas temperature and the core intensity of its formation height; greater height means lower intensity \citep{cauzzietal09, leenaartsetal12}. 

The extended shortward wings of H{$\alpha$} or H{$\beta$} carry signals of unsigned magnetic fields analogous to \ion{Mn}{i} or the {\it{G}}-band \citep{leenaartsetal06a, leenaartsetal06b, leenaartsetal12}.  These line wings obtain contrast enhancement through a reduction of line opacity in magnetic concentrations with locally lesser damping.  However, full-disk spectra face the complication that bright magnetic areas can be dimmed by overlying dark chromospheric filaments.  The emergence, passage, and decay of active solar regions induce variability patterns in Balmer lines \citep{marchenkoetal21}.

\subsection{Balmer-line variability in integrated sunlight}

In the Kitt Peak monitoring program, H$\alpha$ became listed among those lines varying in strength like \ion{Ca}{ii}, but with lower relative amplitudes.  The H$\alpha$ absorption depth was found to drift from $\sim$84.0\% at activity minimum to become slightly shallower ($\sim$\,83.5\%) at activity maximum, a slight filling-in of the line center; \citet{livingstonetal07, livingstonetal10}, further examined by \citet{meunierdelfosse09}.

Balmer lines in the spectrum of the Sun as a star were also evaluated by \citet{criscuolietal23}, using spaceborne radiometers and ground-based spectra.  They identify correlations with network, prominences, filaments and other, finding that both the core and wings contribute to the variability, supporting the view that Balmer line core-to-wing ratios behave more like photospheric than chromospheric indices.  For different Balmer lines, current HARPS-N data were examined by \citet{maldonadoetal19}, including the modulation by solar rotation.  The very cores of the lines were integrated over 0.16 nm, and the wings over about 1 nm.  They confirm certain correlations with the \ion{Ca}{ii} index, more pronounced at times with greater sunspot numbers.  

The Balmer lines display very extended wings that are blended by numerous weaker lines and there is no unique definition of how their precise strengths would be measured.  We adopt the absorption over central portions, referenced to nearby pseudocontinuum levels in their far wings, analogous to previous H$\alpha$ indices, but now covering a wider part of the  profiles, where much outside the very cores must be of photospheric origin.  Our sampling intervals for H$\alpha$ and H$\beta$ are thus much broader than those used by \citet{maldonadoetal19}, and others (Table \ref{table:linelist2}). 

These intervals for H$\alpha$, H$\beta$, and H$\gamma$ are shown in Fig.~\ref{fig:balmer_lines}, and should also be practical to use in foreseen models of magnetically influenced atmospheres (Paper~I was oriented toward metal lines in the photosphere proper, not strong lines in higher layers).  For the desired precision, however, telluric lines touching H$\alpha$ may start to become an issue \citep{dravinsetal15}.  

During the years, there are gradual and systematic changes (Fig.~\ref{fig:ha_hb_hg_abs_mtw}), with H$\alpha$ varying with much greater amplitude than the others.  However, the variations change sign between the successive Balmer lines.  With increased chromospheric activity, H$\alpha$ becomes filled-in (as expected) but the H$\gamma$ absorption instead strengthens, with H$\beta$ fluctuations somewhat indeterminate in between.  This diversity among differently strong lines reminds of that between lines in the \ion{Mg}{i} triplet (Fig.~\ref{fig:mg_b1-b2-b3_abs-mtw}).

For H$\beta$, also its central and less blended core (Fig.\ \ref{fig:balmer_lines}) was measured separately.  A comparison shows the H$\beta$ core to be somewhat less variable than the full line; possibly the wings could be affected by dynamic chromospheric events (Fig.~\ref{fig:balmer_ratios}).  We note that the line-center activity indices in \citet{maldonadoetal19} showed a positive correlation for changes in H$\alpha$, H$\beta$, and H$\gamma$, but their reference wavelengths are different.  Such disparate responses of Balmer and \ion{Mg}{i} lines to magnetic regions should become possible to capture with 3D models, once these are made to extend to also higher atmospheric layers than the photosphere proper.

Especially for H$\beta$, the relation between line strength and \ion{Ca}{ii} emission is not clear  (Fig.~\ref{fig:ha_hb_hg_abs_mtw}).  We note that, in some  fraction of F-G-K stars, complex relationships between \ion{Ca}{ii} and H$\alpha$ chromospheric emission have been seen, including clear anticorrelations \citep{meunieretal22}.  Besides emission from magnetic plages and network, absorbing dark filaments could perhaps be at work \citep{meunierdelfosse09}.  Such filaments could be expected to primarily modify the H$\alpha$ signal, with much smaller effects on the less opaque higher Balmer lines, possibly being a clue to our anticorrelation between H$\alpha$ and H$\gamma$.

\begin{figure}
 \centering
 \includegraphics[width=\hsize]{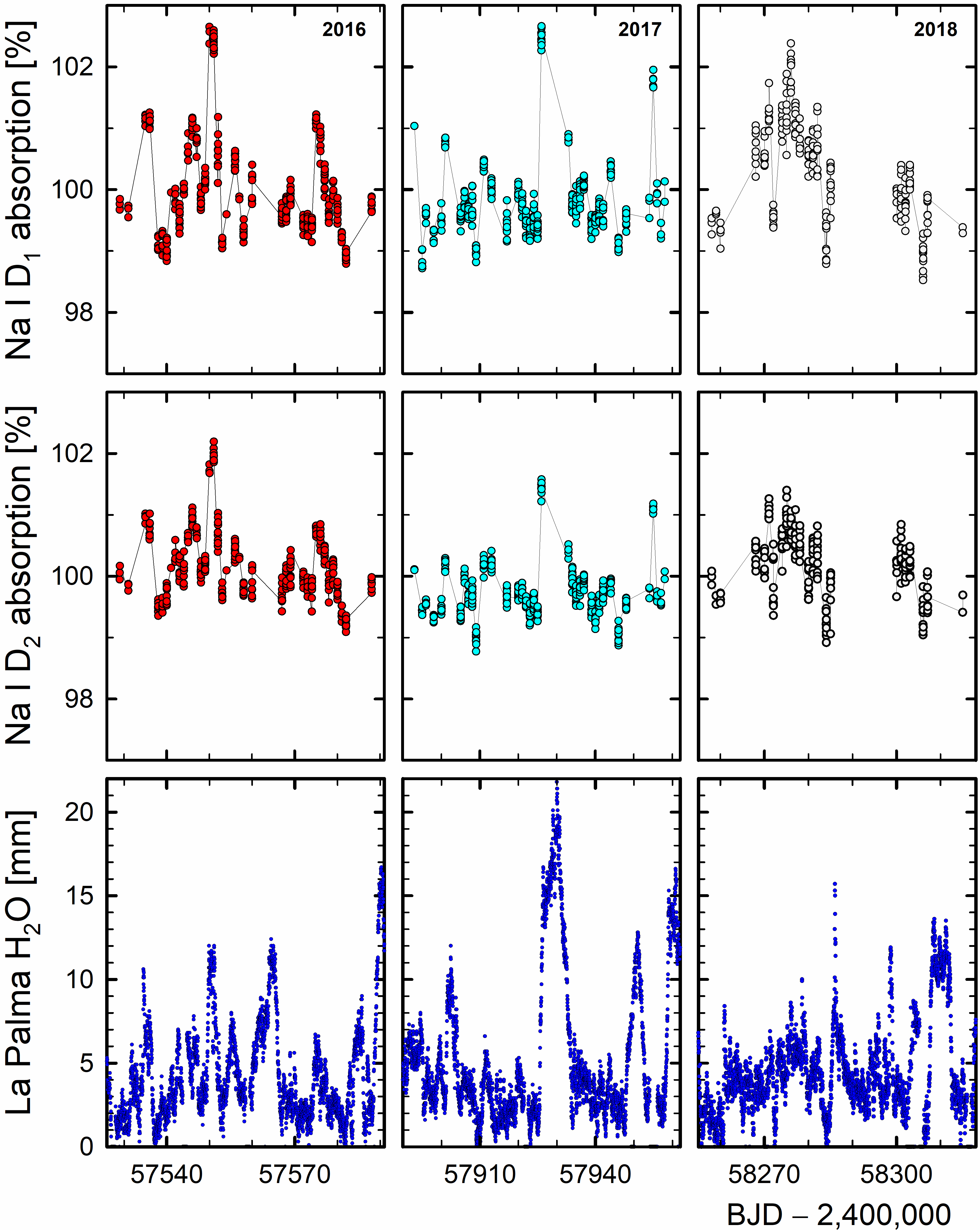}
     \caption{Relative variations of the measured \ion{Na}{i} D$_{1}$ and D$_{2}$ absorption during the seasons of observation.  Bottom: Precipitable water vapor above the Roque de los Muchachos observatory on La Palma.  Time is given as Barycentric Julian Date.  Spectral data from the noisier period in 2018 are omitted (Fig.~\ref{fig:cont_levels}).}  
\label{fig:nai_d1_d2_abs_h2o}
\end{figure}

 \begin{figure*}
 \sidecaption
 \includegraphics[width=11 cm]{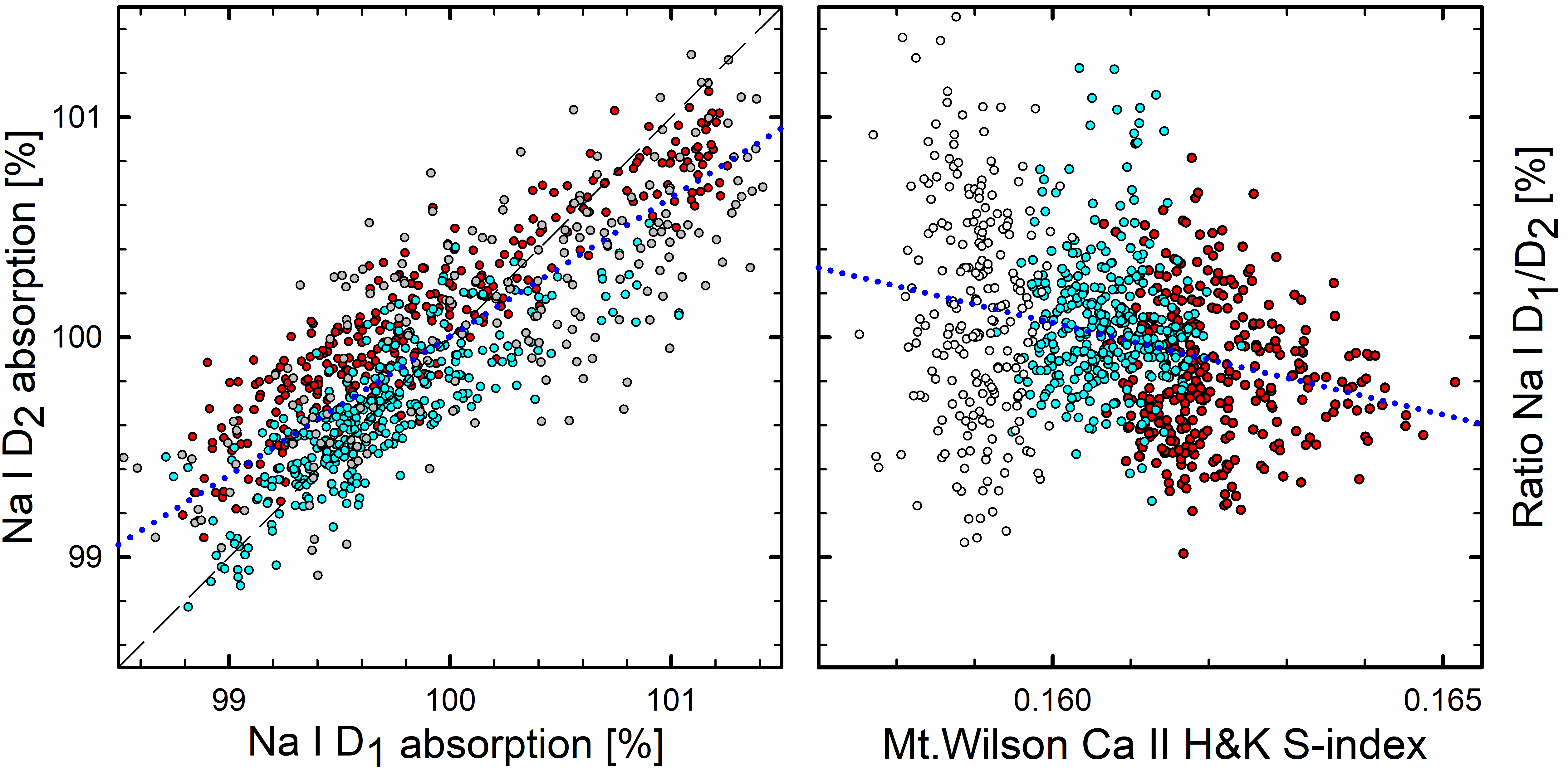}
     \caption{Differential variability of absorption in \ion{Na}{i} D$_{1}$ and D$_{2}$ lines.  The observing seasons of 2016, 2017, and 2018 are marked in red, cyan, and gray.  The identity relation is dashed while linear fits are dotted blue.  The weaker \ion{Na}{i} D$_{1}$ line is more variable than the more saturated D$_{2}$, while their ratio mirrors the chromospheric activity index. }  
\label{fig:nai_d1_vs_d2_abs}
\end{figure*}

\subsection{\ion{Na}{i} D$_1$ and D$_2$ lines}

The \ion{Na}{i} D$_1$ and D$_2$ are two strong lines used as diagnostics for the upper photosphere and the lower chromosphere (Fig.~\ref{fig:chromospheric_lines}).   Usually, the weaker and less blended \ion{Na}{i} D$_1$ is used.  Monochromatic images across the line profiles reveal the atmospheric structures contributing to the local intensity.  The photospheric network is visible in \ion{Na}{i} D$_1$ line wings, but largely vanishes in its core, indicating that one is begins to see the top of granulation proper \citep{ruttenetal11}.  These lines are also used to diagnose exoplanetary atmospheres, but reliable conclusions from the subtle signal then measured differentially to the stellar spectrum, requites a detailed understanding of line formation in 3D and non-LTE \citep{canocchietal24}.  Spatially averaged profiles show a red asymmetry in the core with bisectors in the shape of an inverse C, opposite to that commonly seen in photospheric lines \citep{uitenbroek06}.  Still, the flux in \ion{Na}{i} D$_1$ is largely photospheric.  \citet{leenaartsetal10} find that most of its brightness samples the magnetic network in the photosphere, well below chromospheric heights.  Similar to other stronger lines, magnetic bright points are visible also in \ion{Na}{i} D$_1$ \citep{jessetal10, keysetal13}.

The measured absorption in \ion{Na}{i} D$_{1}$ and D$_{2}$ during the seasons is shown in Fig.\ \ref{fig:nai_d1_d2_abs_h2o}.  Although the amplitudes do not much exceed one percent, the variability is greater than in photospheric lines, and merits a closer examination.  This spectral region is one affected by telluric lines and -- as discussed above -- this may start to become the limiting parameter for ground-based observations.  An example of heavy telluric contamination in \ion{Na}{i} D$_{1}$ and D$_{2}$ was shown by \citet{kjaersgaardetal23}, comparing different removal algorithms.  Although their spectrum -- a winter exposure -- was through airmasses and water vapor levels much higher than ours, they illustrate the perils at these wavelengths.

Fig.\ \ref{fig:nai_d1_d2_abs_h2o} also shows the precipitable water vapor during our data periods as measured above the Roque de los Muchachos observatory on La Palma.  Those atmospheric measurements \citep{castroalmazanetal16} are routinely carried out using GPS instrumentation.  Daily variations can be seen to correlate with apparent changes in line parameters, with times of enhanced water content often coinciding with increased line absorption.  Clearly, such fluctuations in \ion{Na}{i} D$_1$ and D$_2$ do not reflect solar microvariability but rather provide an example of the limits for measurements without detailed modeling of telluric absorption, even for summer observations close to zenith.  No correlations with water vapor were seen for any other among our lines. 

With a reservation for possible differential telluric effects, Fig.~\ref{fig:nai_d1_vs_d2_abs} shows the differential variability between \ion{Na}{i} D$_{1}$ and D$_{2}$.   The amplitudes are greater for the weaker D$_{1}$, which seemingly has more room to strengthen than the more saturated D$_{2}$, also seen in the dependence on the \ion{Ca}{ii}~H\,\&\,K S-index.

\begin{figure*}
 \centering
 \includegraphics[width=\hsize]{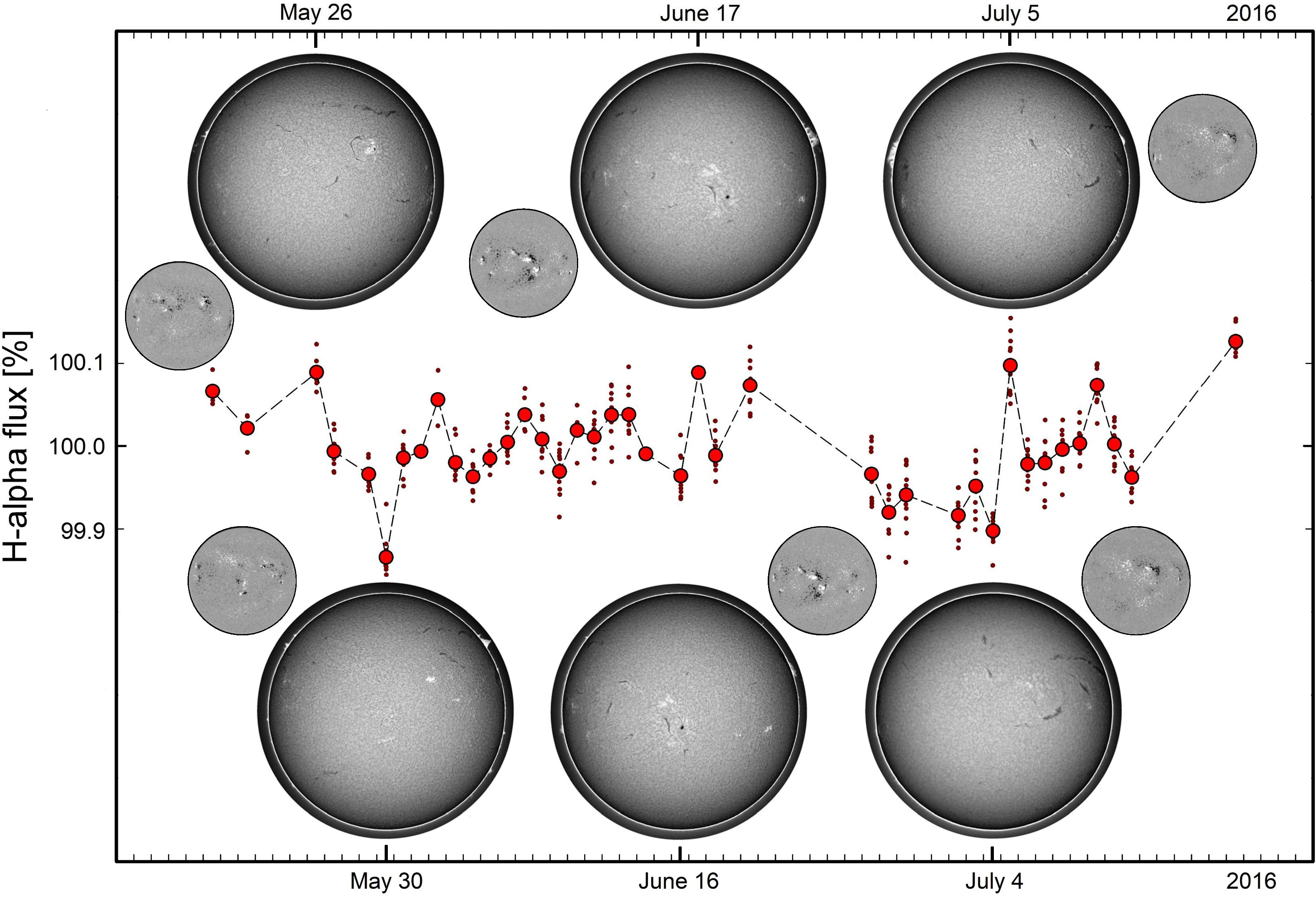}
     \caption{Variations in H$\alpha$ flux (not absorption) measured during the 2016 data period, together with contemporaneous full-disk H$\alpha$ images and (adjacent smaller circles) photospheric magnetograms, at selected labeled dates of local H$\alpha$-flux maxima or minima.  Daily H$\alpha$ flux averages are large red dots, individual exposures are small and black.  The full-disk images were acquired by GONG instruments operated by NISP/NSO/AURA/NSF with contribution from NOAA \citep{gong24}. }  
\label{fig:h-alpha_2016}
\end{figure*}

\subsection{Lines in the near infrared}
 
The infrared Ca triplet lines at $\lambda$\,849.8, $\lambda$\,854.2, and $\lambda$\,866.2 nm reflect activity in stars \citep{busaetal07, foingetal89, huangetal23}, while their synthetic spectra and issues of telluric line contamination are considered by \citet{chmielewski00}.  A comparison between this triplet and \ion{Ca}{ii}~H\,\&\,K is by \citet{martinetal17}.  However, these lines fall longward of the HARPS-N spectral range.

Further into the infrared, the \ion{He}{i} $\lambda$\,1083 mn variability in the Sun as a star was studied by \citet{harvey84, lifeng20, shcherbakovshcherbakova91}, and \citet{shcherbakovetal96}.  Another helium line, \ion{He}{i} D$_{3}$, carries chromospheric signatures in main-sequence stars \citep{dankslambert85}.  In integrated sunlight, such lines are being observed with the NIRPS\footnote{Near InfraRed Planet Searcher} spectrometer at ESO on La Silla \citep{wildietal22}, but the precision is challenged by the need to accurately reduce for the enhanced telluric absorption in the infrared.

\section{Spatially resolved solar activity}

Any variations in integrated sunlight originate from changes in solar surface structures.  Among the lines studied here, H$\alpha$  probably is the only one, where a connection with specific surface phenomena can be identified.  Even if its fluctuations may largely be chromospheric, its behavior may hint at how also photospheric lines are modulated, even if on some lesser order of magnitude. 

Monochromatic H$\alpha$ images are continuously being recorded by the GONG collaboration \citep{gong24}, together with magnetic fields, and other.  Figure \ref{fig:h-alpha_2016} attempts to connect representative H$\alpha$ variability with surface patterns.  For the 2016 summer season, the H$\alpha$ flux (rather than absorption) within its passband (Fig.~\ref{fig:balmer_lines}, Table~\ref{table:linelist2}) is plotted, together with representative full-disk images and magnetograms from particular days exhibiting local flux minima and maxima.  An image examination suggests that instances of higher flux often occur (not surprisingly) when there are larger areas of bright plage on the disk and/or large prominences outside the limb.  Lower flux seems to be common when central disk regions have no distinct plages but more often large dark filaments, a particular signature for H$\alpha$ \citep{meunierdelfosse09}.  However, the correlation of flux levels with magnetic field morphology appears to be weak: although bipolar active regions are shaping the H$\alpha$ structures, those may be either bright plage or dark filaments, not generating any unique correlations.  Such examples demonstrate that fluctuations in the photosphere need to be measured in photospheric lines and, due to the rapid change of physical conditions with height, are unlikely to offer precise proxies in lines formed in higher layers.  Of course, \ion{Ca}{ii}~K chromospheric emission correlates well with the magnetic patterns but even magnetograms recorded in photospheric lines may only represent some particular atmospheric layer since their Zeeman signal is derived from the flanks or wings of the line profiles, which are formed at some height where the fields have already expanded somewhat from their deeper photospheric concentrations.

\section{Conclusions and outlook}

Besides providing very precise wavelength measurements, the long-term stability of radial-velocity spectrometers enables extreme precision stellar spectroscopy.  In observations of the spectrum of the Sun as a star, its microvariability was measured between successive years, with systematically different amplitudes found among different spectral features.  Still, limitations are that current data cover only part of one single activity cycle (cf.\ Fig.~\ref{fig:activitycycles}), and the photometric precision limits the signal for individual lines or narrower spectral segments.  

A pattern with greater line-strength amplitudes for \ion{Fe}{ii} lines in the violet follows the theoretically predicted trends in Paper~I for short-term jittering in radial velocities. The somewhat stronger lines in the green \ion{Mg}{i} triplet show intriguingly variant behavior among themselves, while the semi-forbidden \ion{Mg}{i} $\lambda$\,457.1 nm appears to be stable.  The central parts of the {\it{G}}-band and of H$\beta$ vary with different amplitudes than their full extents.  Such examples highlight that the signatures from different lines or line groups carry dissimilar information and need to be considered separately \citep[e.g.,][and references in Paper~I]{almoullaetal24, demingetal24}. 

To find proxies with the required precision to mitigate radial-velocity fluctuations toward exoEarth detection, it may well be necessary to understand the detailed physical origin of specific line-profile variability and how variations in, e.g., equivalent width relate to wavelength displacements.  Photospheric line changes are seen to correlate with variations in the \ion{Ca}{ii}~H\,\&\,K index, demonstrating the influence of magnetic fields and motivating spectral modeling of the magnetic photosphere.  For different stars, 3D modeling already permits synthesis of hyper-high resolution spectra, revealing details in line profile shapes and wavelength shifts among various classes of spectral features \citep[e.g.,][and Paper~I]{dravinsetal21a}.    

However, the spectral resolution of most radial-velocity instruments may not suffice to test or guide such models since line asymmetries and shifts get significantly degraded already at resolutions $\lambda$/$\Delta\lambda$ $\sim$100,000.  It has been argued that resolutions of such order could be optimal if the criterion is to measure wavelength displacements in the flanks of medium-strength spectral lines using a minimal observing time.  That of course is relevant for large-scale surveys of numerous stars, where observing time is a limit.  However, if the priority is to extract the maximum of physical spectral-line information, and to identify particularly small signals, other criteria apply.

A resolution of 100,000 (corresponding to 3 km\,s$^{-1}$) only provides some three or four independent wavelength points across a typical photospheric line.  While this may provide a value for the line strength and indicate its general sense of asymmetry, further details of the line profile are washed out, as demonstrated in numerous works on both observed and synthetic solar spectra  \citep[e.g.,][]{dravinsetal21b, lohnerbottcheretal19}.  A more satisfying spectral resolution of 270,000, accompanied with a S/N of up to 8,000 was realized for the Sun seen as a star with the PEPSI\footnote{Potsdam Echelle Polarimetric and Spectroscopic Instrument} spectrometer on the LBT\footnote{Large Binocular Telescope} \citep{strassmeieretal18}.  Such numbers would be much preferred for the monitoring of sunlight, as this starts to become adequate to measure actual line shapes.  Such S/N ratios also should enable to identify behavior in individual weaker lines (rather than group averages), and to follow their short-term jittering, segregating from instrumental issues.  Certain monitoring of the solar flux spectrum has actually been carried out with PEPSI \citep{dinevaetal20}, feeding light via a fiber connection from a small solar-disk integration telescope of 13 mm aperture.  A recent upgrade of that telescope has also polarimetric capabilities.  Even higher resolution, and with extremely stable wavelengths, could be achieved from a solar-cycle sequence of FTS spectral atlases, although their specific observing mode and required integration times would not resolve short-term fluctuations \citep{debusetal23, reinersetal16}. 

In the shorter term, a significant improvement should be possible, once the PoET\footnote{Paranal solar ESPRESSO Telescope} solar telescope \citep{leiteetal22} is connected to the ESPRESSO spectrometer of ESO on Paranal \citep{pepeetal21}, enabling sunlight spectroscopy with spectral resolutions $\lambda$/$\Delta\lambda\sim$200,000.  In the longer term, it should be noted that photon fluxes from brighter stars in larger telescopes are fully adequate to record low-noise spectra also at hyper-high resolutions of $\lambda$/$\Delta\lambda\sim$1,000,000, say.  However, the design of corresponding spectrometers remains challenging.  If a cross-dispersed {\'e}chelle remains the design of choice, diffraction-limited adaptive-optics photonic spectrometers will likely be required in order to limit their physical size.  Promising designs are being discussed \citep{bechteretal20, crepp14, jovanovicetal16, jovanovicetal17, plavchanetal19} some of which could be attractive alternatives for later-generation instruments on large, very large, and extremely large telescopes.

\begin{acknowledgements}
{The work by DD is supported by grants from The Royal Physiographic Society of Lund.  We thank Julio A. Castro-Almaz{\'a}n and the Sky Quality Team of the Instituto de Astrof{\'i}sica de Canarias for providing measurements of the precipitable water vapor above the Roque de los Muchachos observatory on La Palma.  For the full-disk solar images, this work utilizes GONG data obtained by the NSO Integrated Synoptic Program, managed by the National Solar Observatory, which is operated by the Association of Universities for Research in Astronomy (AURA), Inc. under a cooperative agreement with the National Science Foundation and with contribution from the National Oceanic and Atmospheric Administration. The GONG network of instruments is hosted by the Big Bear Solar Observatory, High Altitude Observatory, Learmonth Solar Observatory, Udaipur Solar Observatory, Instituto de Astrofísica de Canarias, and Cerro Tololo Interamerican Observatory.  Parts of this paper were completed by DD during a stay as a Scientific Visitor at the European Southern Observatory in Santiago de Chile.  Extensive use was made of NASA’s ADS Bibliographic Services and the arXiv$^{\circledR}$ distribution service. We thank the referee for very knowledgeable and detailed comments. }

\end{acknowledgements}


\begin{appendix}

\section{Issues of photometric stability}

The degree to which microvariability can be measured depends on the stability of the instrument and its calibration.  As described above, line intensities are measured relative to pseudocontinua placed symmetrically about the respective lines.  Their wavelength extents were generally chosen to be wider than the lines to be measured and thus random photometric noise in the lines themselves should not be the limiting factor, but rather the character of possible systematics in the calibration segments. 

\begin{figure}
 \centering
 \includegraphics[width=\hsize]{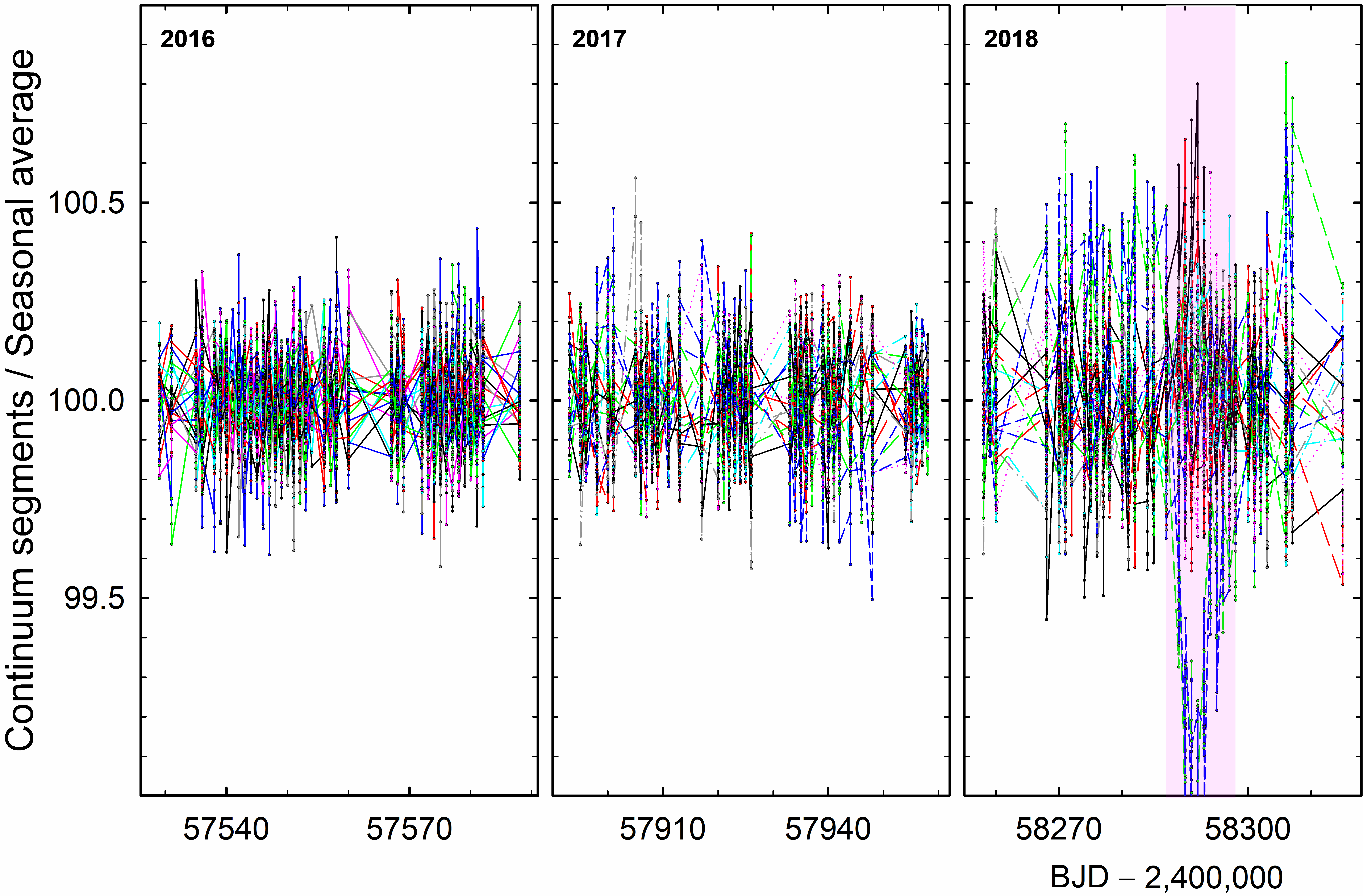}
     \caption{Example of continuum-level stability: relative fluxes in 18 pseudocontinuum segments for the \ion{Fe}{i} 430-445 nm group, each normalized to its seasonal average.  Here, as in other spectral regions, a few weeks of more noisy data appeared in 2018, and were removed from later analyses (shaded in pink). }  
\label{fig:cont_levels}
\end{figure}

A measure of the stability can be obtained by examining the levels of these calibration segments, which are not expected to undergo physical variation.  A representative example is shown in Fig.~\ref{fig:cont_levels}.  This shows the nominal fluxes in the 18  individual pseudocontinuum segments for the \ion{Fe}{i} 430-445 nm group, normalized to the seasonal average for each of them.  The behavior is similar in also other spectral regions, such that the fluctuations are stable around their averages in 2016 and 2017, but are noisier in 2018 with a period of a few weeks with more significant excursions.  

Even if these largest fluctuations may appear striking on the scale of these plots, the amplitudes of even individual continuum segments remain below one percent and would go unnoticed in any more ordinary spectrophotometry.  Still, since the present aim is to push photometry to its limit, data from this noisier period, June 17--28, 2018 (BJD 58,287--58,298), were excluded from all later analyses.  

Except for random photometric noise, it is awkward to precisely quantify the instrumental or observational factors toward limits in photometric precision.  The current analysis uses 1D data reduced from extractions from a 2D {\'e}chelle spectrometer format.  In principle, any ensuing resampling adds noise but resampling cannot be avoided to obtain spectra with both the target lines and a sequence of pseudocontinuum references.  The scatter seen in the Figures throughout the paper serves to illustrate what amplitudes of microvariability that can be identified in spectra from HARPS-like instruments, although it remains awkward to tell what part of the finer scatter is solar and what is instrumental.

\section{Spectral line selection}

\begin{table}[H]
\tiny
\caption{\ion{Fe}{i} and \ion{Fe}{ii} line segments selected for measurements of equivalent widhts.  Short- and long-wavelength limits are shown, together with the number of measured pseudocontinuum points. Wavelength values are those in air. }             
\label{table:1}      
\centering          
\begin{tabular}{c c c c}
\hline      
Spectral feature & Short wave [nm] & Long wave [nm] & \# `Continua' \\ 
\hline   
\\

\ion{Fe}{i} 430-445 a  & 436.5769  &  436.6043  & 18 \\	
\ion{Fe}{i} 430-445 b  &  438.9129  &  438.9381  & 18 \\	
\ion{Fe}{i} 430-445 c  &  442.3732  &  442.3974  & 18 \\	
\ion{Fe}{i} 430-445 d  &  443.2453  &  443.2695  & 18 \\	
\ion{Fe}{i} 430-445 e  &  443.9527  &  444.0013  & 18 \\	
\ion{Fe}{i} 430-445 f  &   444.2697  & 444.3414  & 18 \\	
\ion{Fe}{i} 430-445 g  &  444.5347  &  444.5590  & 18 \\	
\ion{Fe}{i} 430-445 h  &  444.7001  &  444.7257  & 18 \\ 
\\
\ion{Fe}{i} 520-535 a  &  524.2205  &  524.4098  & 8 \\	
\ion{Fe}{i} 520-535 b  &  524.6638  &  524.8188  & 8 \\	
\ion{Fe}{i} 520-535 c  &  524.9911  &  525.1089  & 8 \\	
\ion{Fe}{i} 520-535 d  &  525.2812  & 525.3790  & 8  \\	
\\
\ion{Fe}{i} 670-685 a  &  672.5100  & 672.6995  &  16  \\	
\ion{Fe}{i} 670-685 b  &  674.9813  &  675.0496  &  16  \\
\ion{Fe}{i} 670-685 c  &  675.2305  &  675.3007  &  16  \\
\ion{Fe}{i} 670-685 d  &  676.7412  &  677.2800  &  16  \\
\ion{Fe}{i} 670-685 e  &  678.6616  &  678.7284  &  16  \\
\ion{Fe}{i} 670-685 f  &  680.6582  &  680.7103  &  16  \\
\ion{Fe}{i} 670-685 g  &  680.9915  &  681.0586  &  16  \\
\ion{Fe}{i} 670-685 h  &  683.6716  &  684.3994  &  16  \\
\ion{Fe}{i} 670-685 i  &   685.4016  &  685.5929  &  16  \\
\ion{Fe}{i} 670-685 j  &  685.6998  &  685.8442  &  16  \\
\\
\ion{Fe}{ii} 435-475 a  &  441.3471  & 441.3725  &  47  \\
\ion{Fe}{ii} 435-475 b  &  441.6671  &  441.7070  &  47  \\
\ion{Fe}{ii} 435-475 c  &  449.1253  &  449.1560  &  47  \\
\ion{Fe}{ii} 435-475 d  &  450.8127  &  450.8472   &  47  \\
\ion{Fe}{ii} 435-475 e  &  451.5013   &  451.5741  &  47  \\
\ion{Fe}{ii} 435-475 f  &   452.0091  &  452.0388  &  47  \\
\ion{Fe}{ii} 435-475 g  &  454.1394   &  454.1730  &  47  \\
\ion{Fe}{ii} 435-475 h  &  457.6147  &  457.6460  &  47  \\
\ion{Fe}{ii} 435-475 i  &  458.2660  &  458.3036  &  47  \\
\ion{Fe}{ii} 435-475 j  &  462.0217  & 462.0735  &  47  \\
\ion{Fe}{ii} 435-475 k  &  465.6846  &  465.7088  &  47  \\
\ion{Fe}{ii} 435-475 l  &  466.6664  &  466.6856  &  47  \\
\ion{Fe}{ii} 435-475 m  &  467.0010  & 467.0291  &  47  \\
\ion{Fe}{ii} 435-475 n  &  473.1288  &  473.1663  &  47  \\
\\
\ion{Fe}{i} @ {\it{G}}-band a  &  426.4083  &  426.4375  &  32  \\
\ion{Fe}{i} @ {\it{G}}-band b  &  426.5786  &  426.6066  &  32  \\
\ion{Fe}{i} @ {\it{G}}-band c  &  426.6836  &  426.7105  &  32  \\
\ion{Fe}{i} @ {\it{G}}-band d  &  428.6313  & 428.6746  &  32  \\
\ion{Fe}{i} @ {\it{G}}-band e  &  429.1334  &  429.1651  &  32  \\
\ion{Fe}{i} @ {\it{G}}-band f  &  431.8475  &  431.8889  &  32  \\
\ion{Fe}{i} @ {\it{G}}-band g  &  432.7000  &  432.7296  &  32  \\
\ion{Fe}{i} @ {\it{G}}-band h  & 433.6906  &  433.7179  &  32  \\
\ion{Fe}{i} @ {\it{G}}-band i  & 434.7096  &  434.7370  &  32  \\
\ion{Fe}{i} @ {\it{G}}-band j  &  434.8809  &  434.9106  &  32  \\
\hline       
\end{tabular}
\label{table:linelist1}
\end{table}

\begin{table}
\tiny
\caption{Non-Fe line segments selected for measurements of equivalent widhts. }             
\label{table:1}      
\centering          
\begin{tabular}{c c c c}
\hline      
Spectral feature & Short wave [nm] & Long wave [nm] & \# `Continua' \\ 
\hline   
\\
\ion{Mg}{i} 457.1 & 457.0955 & 457.1258 & 4 \\
\ion{Mg}{i} b{$_1$} 518.3 & 516.6461 & 516.8454 & 8 \\
\ion{Mg}{i} b{$_2$} 517.2 & 517.1820 & 517.3617 & 8 \\  
\ion{Mg}{i} b{$_3$} 516.7 & 518.1477 & 518.5716 & 8 \\
\\
\ion{Mn}{i} hyperfine a  &  511.7797  &  511.8049   &  2  \\
\ion{Mn}{i} hyperfine b  &  525.5212  & 525.5457  &  3  \\
\ion{Mn}{i} hyperfine c  &  537.7428  &  537.7811  &  2  \\
\ion{Mn}{i} hyperfine d  &  539.4473  &  539.4886  &  2  \\
\ion{Mn}{i} hyperfine e  &  539.9271  &  539.9699   &  4  \\
\ion{Mn}{i} hyperfine f  &   540.7192  &  540.7577   &  3  \\
\ion{Mn}{i} hyperfine g  &  541.3497  &  541.3852  &  2  \\
\ion{Mn}{i} hyperfine h  &  542.0105  &  542.0579  &  3  \\
\ion{Mn}{i} hyperfine i  &  543.2379   &  543.2736  &  3  \\
\ion{Mn}{i} hyperfine j  &   545.7265  & 545.7638  &  2  \\
\ion{Mn}{i} hyperfine k &   547.0477  &  547.0881  &  3  \\
\ion{Mn}{i} hyperfine l  &   551.6577  &  551.6954  &  3  \\
\ion{Mn}{i} hyperfine m  &  553.7530  &  553.8015  &  5  \\
\ion{Mn}{i} hyperfine n  &   601.3277  &  601.3754  &  2  \\
\ion{Mn}{i} hyperfine o  &   602.1523  &  602.2066  &  2  \\
\ion{Mn}{i} hyperfine p  &   644.0776  &   644.1111  &  2  \\
\\
{\it{G}}-band full  &  429.8359  &  431.3281  &  32  \\
{\it{G}}-band core  &  430.7056   &  430.8753  &  32  \\
\\
{H$\alpha$} & 655.6222 & 656.8571 & 2 \\
{H$\beta$ full} & 485.8459 & 486.3458 & 6 \\
{H$\beta$ core} & 486.0027 & 486.2434 & 6 \\
{H$\gamma$} & 433.9861 & 434.1167 & 2 \\
\\
\ion{Na}{i} D{$_1$} 589.59 & 589.4495 & 589.7011 & 5 \\
\ion{Na}{i} D{$_2$} 588.99 & 588.7808 & 589.1046 & 5 \\
\hline       
\end{tabular}
\label{table:linelist2}
\end{table}

\end{appendix}

\end{document}